\documentclass{article}
\usepackage[a4paper,top=1in,bottom=1.25in]{geometry}
\usepackage[intlimits]{amsmath}
\usepackage[pdftex]{graphicx}
\usepackage[multidot,space]{grffile}
\usepackage{longtable}
\usepackage{ifthen}
\usepackage[english]{varioref}
\usepackage{hyperref}
\usepackage{refcount}
\usepackage{gensymb}
\usepackage{amssymb}
\usepackage{wasysym}
\usepackage{textcomp}
\usepackage{float}
\usepackage{fancyhdr}
\usepackage{extramarks}
\usepackage{ulem}
\usepackage{caption}
\usepackage{subfig}
\usepackage{wrapfig}
\usepackage{footnote}
\usepackage{multirow}
\usepackage{tabulary}
\usepackage[procnames]{listings}
\usepackage[usenames,dvipsnames,svgnames,table]{xcolor}
\usepackage{numprint}
\usepackage{natbib}
\usepackage[toc]{appendix}
\usepackage[nottoc,numbib]{tocbibind}
\usepackage[english]{babel}

\definecolor{text}{HTML}{000000}
\definecolor{keyword}{HTML}{0000FF}
\definecolor{builtin}{HTML}{900090}
\definecolor{definition}{HTML}{000000} 
\definecolor{comment}{HTML}{ADADAD} 
\definecolor{string}{HTML}{00AA00}
\definecolor{number}{HTML}{800000}
\definecolor{instance}{HTML}{924900} 
\definecolor{linenumber}{HTML}{ADADAD} 
\makeatletter
\newif\iffirstchar\firstchartrue
\newif\ifstartedbyadigit

\newcommand\ProcessLetter
{%
	\ifnum\lst@mode=\lst@Pmode%
		\iffirstchar%
				\global\startedbyadigitfalse%
			\fi
			\global\firstcharfalse%
		\fi
}
\newcommand\ProcessDigit
{%
	\ifnum\lst@mode=\lst@Pmode%
		\iffirstchar%
				\global\startedbyadigittrue%
			\fi
			\global\firstcharfalse%
  \fi
}
\lst@AddToHook{Output}%
{%
	\ifstartedbyadigit%
		\def\lst@thestyle{\color{number}}%
	\fi
	\global\firstchartrue%
	\global\startedbyadigitfalse%
}
\newtoks\python@toks
\python@toks={language=Python,frame=single,commentstyle=\itshape\color{comment},title=\lstname,basicstyle=\footnotesize\rmfamily\color{text},keywordstyle=[1]\color{keyword},keywordstyle=[2]\color{builtin},keywordstyle=[3]\itshape\color{instance},stringstyle=\color{string},identifierstyle=\color{text},morestring=[d]{"""},showspaces=false,showstringspaces=false,numbers=left,numbersep=5pt,numberstyle=\tiny\color{linenumber},breaklines=true,upquote=true,procnamekeys={def,class},procnamestyle=\bfseries\color{definition},morekeywords=[1]{as},morekeywords=[2]{True,False},morekeywords=[3]{self},alsoletter={0123456789},SelectCharTable=%
}
\def\add@savedef#1#2{%
  \begingroup\lccode`?=#1\relax
  \lowercase{\endgroup
  \edef\@temp{%
    \noexpand\lst@DefSaveDef{\number#1}%
    \expandafter\noexpand\csname lsts@?\endcsname{%
      \expandafter\noexpand\csname lsts@?\endcsname\noexpand#2}%
  }}%
  \python@toks=\expandafter{\the\expandafter\python@toks\@temp}%
}
\count@=`0
\loop
  \add@savedef\count@\ProcessDigit
  \ifnum\count@<`9
  \advance\count@\@ne
\repeat
\count@=`A
\loop
  \add@savedef\count@\ProcessLetter
  \ifnum\count@<`Z
  \advance\count@\@ne
\repeat
\count@=`a
\loop
  \add@savedef\count@\ProcessLetter
  \ifnum\count@<`z
  \advance\count@\@ne
\repeat
\begingroup\edef\x{\endgroup
  \noexpand\lstdefinestyle{defaultpython}{\the\python@toks}
}\x
\makeatother

\makeatletter
\newcommand*\TY@cap@gobble[2][]{\\}
\def\ltabulary{%
    \def\caption{
        \@ifstar\TY@cap@gobble\TY@cap@gobble}
    \def\endfirsthead{\\}%
    \def\endhead{\\}%
    \def\endfoot{\\}%
    \def\endlastfoot{\\}%
    \def\tabulary{%
        \def\TY@final{%
    \def\endfirsthead{\LT@end@hd@ft\LT@firsthead}%
    \def\endhead{\LT@end@hd@ft\LT@head}%
    \def\endfoot{\LT@end@hd@ft\LT@foot}%
    \def\endlastfoot{\LT@end@hd@ft\LT@lastfoot}%
    \longtable}%
        \let\endTY@final\endlongtable
        \TY@tabular}%
    \dimen@\columnwidth
    \advance\dimen@-\LTleft
    \advance\dimen@-\LTright
    \tabulary\dimen@}

\makeatother

\makesavenoteenv{tabular}
\makesavenoteenv{environmentname}
\numberwithin{equation}{section}
\labelformat{chapter}{Chap.~#1}
\labelformat{section}{Sec.~#1}
\labelformat{appendix}{App.~#1}
\labelformat{subsection}{Sec.~#1}
\labelformat{subsubsection}{Sec.~#1}
\labelformat{figure}{Fig.~#1}
\labelformat{subfigure}{Fig.~\thefigure #1}
\labelformat{table}{Tab.~#1}
\labelformat{equation}{Eq.~(#1)}

\makeatletter
\let\latex@@vref\vref
\let\latex@@ref\ref
\newcounter{pagegapthreshhold}
\newcounter{tmpcntr}
\renewcommand{\vref}[1]{%
	\setcounter{tmpcntr}{\value{page}}%
	\addtocounter{tmpcntr}{-\getpagerefnumber{#1}}%
	\ifnum\value{tmpcntr} < 0 %
		\setcounter{tmpcntr}{\numexpr -1*\value{tmpcntr}}%
	\fi%
	\ifnum\value{tmpcntr} > \value{pagegapthreshhold}
		\latex@@vref*{#1}%
	\else%
		\latex@@ref{#1}%
	\fi%
}
\let\latex@@vrefrange\vrefrange
\renewcommand{\vrefrange}[2]{%
	\setcounter{tmpcntr}{\value{page}}%
	\addtocounter{tmpcntr}{-\getpagerefnumber{#1}}%
	\ifnum\value{tmpcntr} < 0 %
		\setcounter{tmpcntr}{\numexpr -1*\value{tmpcntr}}%
	\fi%
	\ifnum\value{tmpcntr} > \value{pagegapthreshhold}
		\latex@@vrefrange{#1}{#2}%
	\else%
		\latex@@ref{#1} to \latex@@ref{#2}%
	\fi%
}
\makeatother

\newcommand{\textcite}{\citet}
\newcommand{\autocite}{\citep}

\noappendicestocpagenum
\graphicspath{{Images/}}
\newcommand{\HRule}{\rule{\linewidth}{0.5mm}}
\begin{document}

\newgeometry{a4paper,top=1in,bottom=1.25in,left=1.25in,right=1.25in}
\begin{titlepage}
\begin{center}
\includegraphics[width=0.3\textwidth]{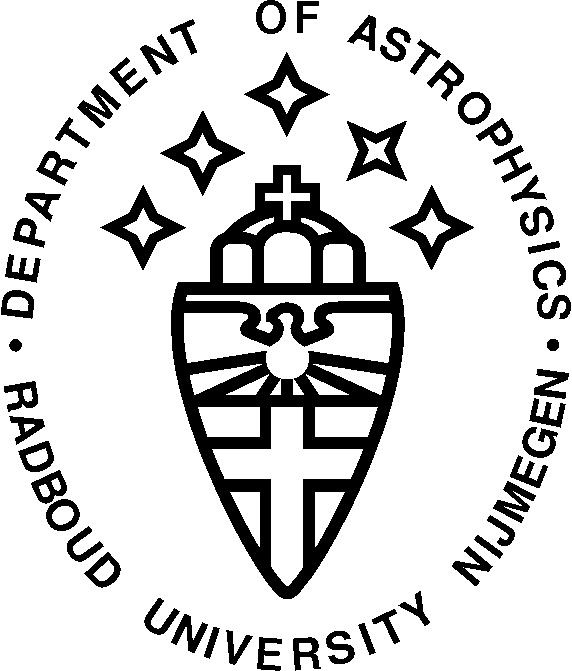}~\\[1cm]
\textsc{\huge{Radboud University}}\\[2.5cm]
\textsc{\LARGE{Master Thesis}}\\[0.5cm]
\HRule\\[0.4cm]
\huge{\textbf{IMAGINE: Testing a Bayesian pipeline for Galactic Magnetic Field model optimization}}
\HRule\\[1.5cm]
\noindent
\begin{minipage}[t]{0.4\textwidth}
\begin{flushleft} \large
\emph{Author:}\\
Ellert \textsc{van der Velden}\\
~\\
\emph{Department:}\\
Institute for Mathematics, Astrophysics and Particle Physics\\
Radboud University, Nijmegen, The Netherlands
\end{flushleft}
\end{minipage}%
\begin{minipage}[t]{0.4\textwidth}
\begin{flushright} \large
\emph{Supervisors:}\\
Dr.~Marijke \textsc{Haverkorn}\footnotemark[1]\\
Dr.~J\"org \textsc{Rachen}\footnotemark[1]\\ 
~\\
\emph{In collaboration with:}\\
Theo \textsc{Steininger}\footnotemark[2]
\end{flushright}
\end{minipage}
\vfill
\Large{July 4, 2017}
\footnotetext[1]{Department of Astrophysics/IMAPP, Radboud University, Nijmegen, The Netherlands}
\footnotetext[2]{Max-Planck-Institut f\"ur Astrophysik, Max-Planck-Gesellschaft, Garching, Germany}
\end{center}
\end{titlepage}
\restoregeometry

\thispagestyle{empty}
\texttt{ }\\
\newpage

\thispagestyle{empty}
\tableofcontents
\newpage

\thispagestyle{empty}
\texttt{ }\\
\newpage

\pagestyle{fancy}
\setcounter{page}{5}
\begin{center}
\textbf{Note:}
\end{center}
This work contains the details and results of my master's project on testing the IMAGINE pipeline for Galactic magnetic field estimation.
The project was carried out from early 2016 to early 2017.
For it, an unpublished early development version of the IMAGINE pipeline was tested and debugged.
The thesis reports about the kind of difficulties faced when dealing with high dimensional complex parametric Galactic magnetic field models.
It was found that such models require extra caution to allow for dependencies between parameters and model implementation errors, which need to be taken into account when performing a Bayesian analysis.
These findings, reported here in this thesis, helped to resolve such issues in the later, now published version of the IMAGINE pipeline.
The thesis therefore documents the genesis of the pipeline and lessons learned during this process.
This document contains original text of the master thesis for reference.
Parts of its content therefore do not reflect the current state of the IMAGINE pipeline.

\begin{abstract}
Obtaining a reliable model of the Galactic magnetic field (GMF) is something many fields in astrophysics long for.
Recently, particular interest in this has been shown by the attempt to understand the origin of ultra-high-energy cosmic rays (UHECRs); charged particles with extremely high energies for which detailed knowledge of the GMF is required to trace them back to their sources.
The development of a GMF model often involves the optimization of a given parametrization by comparing it to data sets.
These data sets can include rotation measures from polarized sources or Galactic synchrotron radiation maps.
However, how reliable are these optimizations and how valid are these models?

The IMAGINE pipeline is designed to answer these questions by using Bayesian statistics.
This allows the pipeline to compare different model approaches with each other while also taking care of the growing amount of available data sets.
During my masters project, I have tested the first version of IMAGINE for one of the most used GMF models, the "JF12 model".
These tests were focused on finding the best sampling method to explore the $30$ dimensions of this model.

In this thesis, I will report on the work that I have done on testing the IMAGINE pipeline and explain the challenges and difficulties of doing so.
The main part of these tests involved analyzing the behavior of the posterior probability distribution functions of the various parameters in the JF12 model.
Doing so has shown that many different numerical errors can pop up when using models, some that can be removed or circumvented and some that are simply hard-coded.
To ensure compatibility with future GMF models, this outcome forces us to use a nested sampler in the IMAGINE pipeline.
The carried out tests have also pointed out various other weak-points in the pipeline, including dependencies between parameters, not taking care of unphysical model results and insufficient control over the modules in the pipeline.
\end{abstract}

\section{Introduction}
\label{sec:Introduction}
Looking up to the sky in the middle of a starry night, watching the twinkling of the stars,\footnote{Something most observational astronomers dislike due to the effects of seeing.} and maybe even recognizing the Milky Way, the world seems so peaceful.
One would not expect this beautiful sky to bombard our world with millions of particles every second.
High-energy photons, protons, electrons and atomic nuclei hit the atmosphere of the Earth in large quantities however, usually producing phenomena called air showers.
The latter three are usually referred to as \textit{cosmic rays} (CRs) and can be found within almost every energy range.
Cosmic rays with the highest of energies, commonly known as \textit{ultra-high-energy cosmic rays} (UHECRs), usually find themselves in the $\mathrm{EeV}$ energy range.
These cosmic rays are special, because their energies are so high, that they can only be produced in a handful of different high-energetic accelerators.
These accelerators are thought to include the jets of active galactic nuclei or gamma-ray bursts, which are not very common astrophysical objects.

It is not known if the mentioned objects can accelerate particles, which is why UHECRs offer an unique opportunity to study these properties.
For example, simply backtracking the path of a detected UHECR could possibly reveal the position of such object.
However, since cosmic rays are usually either atomic nuclei (stripped of their electrons) or solitary protons\footnote{Protons are in principle the atomic nuclei of hydrogen atoms, but exist in such large quantities that they are stated specifically.}, their paths are affected by magnetic fields.
As will be explained in \ref{subsec:GMFs}, our Milky Way has a magnetic field of its own, deflecting cosmic rays as they arrive.
Therefore, if one wants to backtrack the trajectory of a UHECR, a reliable model for the Galactic magnetic field is required.

Many GMF models already exist, the one more sophisticated than the other.
However, the problem with most models is that they are capable of explaining the data set they were made for, but little else.
Models also tend to use different sets of assumptions.
This makes models incompatible with each other, since it is impossible to compare the different assumption sets with each other in this way.
It would therefore be great if something existed that is actually capable of assigning a value to the used assumptions and data, that represents its weight and quality in the model.
This would then allow one to answer the question which GMF model is currently the best.

This is where the work of this thesis comes in: the IMAGINE pipeline.
The IMAGINE pipeline tries to give an answer to this question, by numerically optimizing GMF models to the available data.
Instead of using the GMF models to explain the data sets they were made for, IMAGINE changes the purpose of every model to predict all the available data by using Bayesian statistics.
The model that can do this the best can then be considered currently the best model for the Galactic magnetic field.

This thesis is structured in the following way:
\ref{sec:Theoretical Background} describes the different aspects and subjects that are required in order to make such a pipeline.
\ref{subsec:GMFs} discusses the Galactic magnetic field and its various aspects, while \ref{subsec:GMF Models} gives an overview of the various features in GMF models and describes the "JF12 model" that is used in this thesis.
\ref{subsec:Probability Theory} introduces Bayesian statistics which, together with Markov chain Monte Carlo methods in \ref{subsec:MCMC and Sampling Methods}, is the core ingredient in the IMAGINE pipeline (\ref{sec:IMAGINE Pipeline}).
The obtained information will then be used for a series of tests in \ref{sec:Testing}.

\newpage
\section{Theoretical Background}
\label{sec:Theoretical Background}
\subsection{Galactic Magnetic Fields}
\label{subsec:GMFs}
Following \textcite{txt:Haverkorn}, one can see the Milky Way as a dynamic environment, consisting for the most part out of plasma, such as stars, jets, H\textsc{II} regions and supernova remnants (SNRs), with interstellar medium (ISM) in between.
Since plasma usually generates magnetic fields, it is not surprising that magnetic fields are found indirectly everywhere in the galaxy.

Galactic magnetic fields are traditionally divided into large-scale (following the global structure of a galaxy) and small-scale (typically associated with turbulence).
In the Milky Way, the small-scale field is typically a factor of a few stronger than the large-scale field.

Fully mapping out the strength, direction and structure of the Galactic magnetic field is extremely difficult.
This is because it is made up by a combination of a large-scale and multiple small-scale components.
In addition to that, the available observational methods (\textit{tracers}) can only measure either one component of the magnetic field\footnote{Strength or direction, parallel or perpendicular to the line-of-sight} and/or in one particular medium\footnote{Ionized gas, dense cold gas, dense dust, diffuse dust, etc.}.
The components of the magnetic field that can be measured are integrated over the line-of-sight (the imaginary line drawn between the observer and the observed source), making these tracers two-dimensional at best.
Faraday rotation for example, is only a measure for the average strength of the magnetic field parallel to the line-of-sight $\left\langle B_{\parallel}\right\rangle$ in an ionized medium.
Synchrotron radiation at the other hand, is a measure for the strength of the magnetic field perpendicular to the line-of-sight $B_{\perp}$. 

However, the biggest problem in determining the Galactic magnetic field stems from the vantage point we are forced to use: Earth.
Since the Milky Way has a non-trivial three-dimensional structure, its magnetic field structure is non-trivial as well in three dimensions.
Creating a three-dimensional picture would necessitate three-dimensional tracers, so therefore one or two-dimensional tracers are insufficient.
This necessitates the usage of many assumptions and models about the magnetic field, as well as about the thermal (Faraday) and relativistic (synchrotron) electron distributions. 

\subsubsection{Large-scale \& Small-scale Fields}
\label{subsubsec:Large-scale and Small-scale Fields}
The terminology of large-scale and small-scale Galactic magnetic fields in the literature can differ a lot, due to different authors using different terminology for the same magnetic field configurations or vice versa.
Therefore, an overview over these different magnetic field configurations will be given along with their commonly used names, of which one will be used throughout the rest of this thesis.\footnote{Please refer to \ref{app:Abbreviations/Terminology} for an overview of all used abbreviations and terminology in this thesis.}

As said before, Galactic magnetic fields are usually divided up into large-scale fields and small-scale fields.
The \textit{large-scale} field (also commonly called \underline{regular}, uniform or coherent\footnote{\underline{Regular} will be used throughout this thesis.}) is the component of the magnetic field that is coherent on length scales of the order of a galaxy and usually assumed to follow a pattern, like following the spiral arms of a galaxy.
There are multiple possible explanations for what exactly generates the large-scale field, including a Galactic dynamo \autocite{txt:Shukurov}.

\textit{Small-scale} fields (called \underline{random}, tangled or turbulent) describe the magnetic field components connected to the turbulent ISM \autocite{txt:Ferriere}.
Small-scale fields show large variations in strength, direction and orientation on scales of the order of several parsec to a hundred parsec \autocite{txt:Haverkorn_Brown}, but usually do not extend to scales larger than that.

However, the small-scale field component can be divided into two different components.
One is a true random small-scale field called \underline{isotropic random}\footnote{A random field is sometimes also called isotropic random. In this thesis, a \textit{random field} always means the small-scale field, not specifically one of the two variants.}, which shows variations in magnetic field strength, direction and orientation.
The other is a semi-random small-scale field called \underline{anisotropic random}, ordered random or striated; which has variations in magnetic field strength and direction on small scales, but not in magnetic field orientation.
Such a field can arise when a turbulent field structure is compressed into two dimensions.
This can happen for example in SNR shocks, spiral arm density waves or Galactic shear.

\begin{figure}[htb!]
\begin{center}
	\subfloat[Regular]{\includegraphics[width=0.3\textwidth]{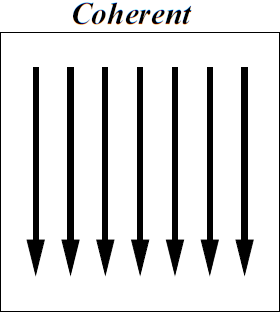}}
	\subfloat[Anisotropic Random]{\includegraphics[width=0.3\textwidth]{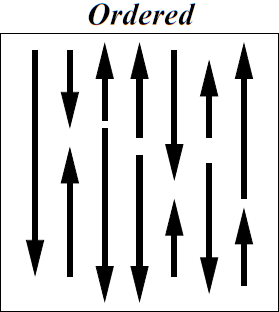}}
	\subfloat[Isotropic Random]{\includegraphics[width=0.3\textwidth]{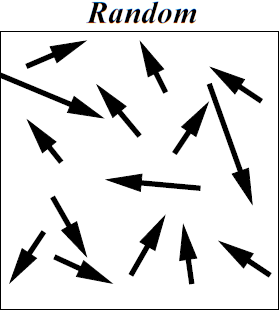}}
\caption{Schematic drawing of the three different magnetic field components.
Note that the regular field is called \textit{coherent}, the anisotropic random field is called \textit{ordered} and the isotropic random field is called \textit{random}.
Also note that these three components have been discussed in a different order in the text.
Reproduced from \textcite{txt:Jaffe}.}
\label{fig:Components Magnetic Fields Individual}
\end{center}
\end{figure}

A schematic drawing of these three different components of the magnetic field can be found in \vref{fig:Components Magnetic Fields Individual}.

\subsubsection{Faraday Rotation}
\label{subsubsec:Faraday Rotation}
\begin{figure}[htb!]
\begin{center}
	\includegraphics[width=\textwidth]{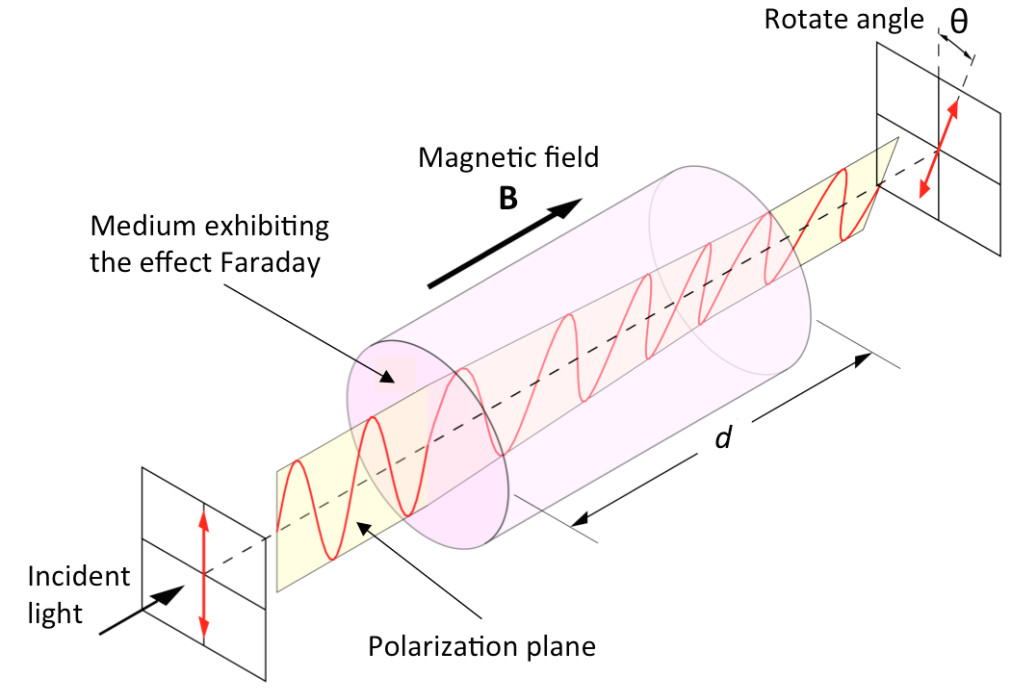}
	\caption{A schematic drawing of the polarization rotation due to the Faraday effect.}
	\label{fig:Faraday}
\end{center}
\end{figure}

Most of the knowledge of the geometry of the large-scale magnetic field comes from studies using the effect of Faraday rotation.
Following the description given in \textcite{txt:Mao}, the Faraday rotation can be described as a double refraction effect when linearly polarized light travels through a magnetized (ionized) medium.
A schematic drawing of this can be seen in \vref{fig:Faraday}.
The polarization angle of the Faraday rotation is given by
\begin{eqnarray}
\label{eq:Faraday Rotation}
\theta&=&\theta_0+\mathrm{RM}*\lambda^2,
\end{eqnarray}
with $\theta$ the polarization angle at detection, $\theta_0$ the polarization angle at emission, RM the rotation measure\footnote{The term RM will be used for both \textit{rotation measure} and \textit{Faraday depth}, because the distinction is not relevant for this thesis. A more in-depth discussion of and distinction between RM and Faraday depth can be found in \textcite{txt:Beck}.} and $\lambda$ the wavelength of the light ray. 
Here, the initial polarization angle $\theta_0$ is the same for all light rays coming from the same object, assuming that only a constant magnetic field is present.
The constant RM depends on the strength of the magnetic field and the density of thermal electrons along the line-of-sight.
The RM is independent of the wavelength of the light ray.
Using this information together with measured wavelengths and polarization angles, a value for the RM can be calculated for a specific source. 

The RM itself, corresponding to a position at a distance $l_0$ from an observer, is given by a line-of-sight integral,
\begin{eqnarray}
\label{eq:RM}
\mathrm{RM}&=&\frac{e^3}{2\pi m_e^2c^4}\int_{l_0}^0 n_e(l)B_{\parallel}(l)dl,
\end{eqnarray}
which can also be written as 
\begin{eqnarray}
\label{eq:RMcgs}
\frac{\mathrm{RM}}{\mathrm{rad\ m^{-2}}}&=&0.812\int_{l_0}^0 \frac{n_e(l)}{\mathrm{cm^{-3}}}\frac{B_{\parallel}(l)}{\mathrm{\mu G}}\frac{dl}{\mathrm{pc}},
\end{eqnarray}
with $n_e(l)$ the thermal electron density at distance $l$, $B_{\parallel}(l)$ the strength of the parallel magnetic field at distance $l$ and $l$ the distance from the observer in the direction of the source.
By convention, the RM is positive (negative) when the magnetic field is moving towards (away from) the observer.

When observing this effect for extragalactic sources, the RM will not contain contributions from the Milky Way alone, but rather from every single position along the line-of-sight to the source.
It would therefore be better to only observe Galactic sources in order to study the large-scale Galactic magnetic field.
Since the Faraday rotation is observed best when using bright sources with a large range of different wavelengths, pulsars are commonly used as Galactic sources.
However, pulsars have a drawback: There are only a few pulsars ($\sim 2000$ with reasonably accurate RMs), which are mostly concentrated in and close to the Galactic plane.
This means that if one wants to gain information about the small-scale structure or information at high Galactic latitudes, extragalactic sources are required.
An example of this is given by \textcite{txt:Oppermann}; an RM map based on $41,330$ measurements of the Faraday rotation of extragalactic point sources.

\subsubsection{Synchrotron Radiation}
\label{subsubsec:Synchrotron}
Synchrotron radiation is caused by the acceleration of relativistic electrons in a magnetic field.
This electromagnetic radiation is then emitted radially to the acceleration, partially in a polarized form.
This polarization is commonly written in terms of the so-called \textit{Stokes parameters}.

The Stokes parameters are a set of four different values that can describe the polarization state of photons.
Although an individual photon is always polarized, the Stokes parameters are used more for the description of the polarization of the total photon beam.
Multiple naming conventions for these four values are used throughout the literature, but the letters $I$, $Q$, $U$ and $V$ are the most common.

\begin{figure}[htb!]
\begin{center}
	\includegraphics[width=0.75\textwidth]{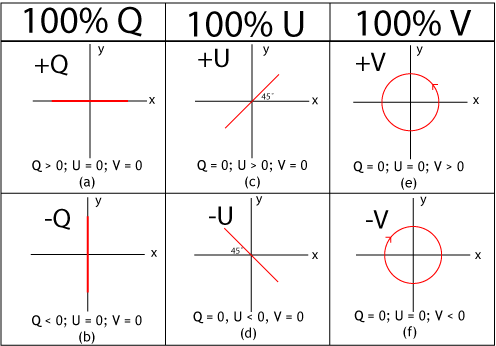}
\caption{Schematic drawing of the three Stokes parameters.
$Q$ describes the linearly polarized intensity in the horizontal or vertical direction.
$U$ describes the same, but rotated over an angle of $45\degree$.
$V$ describes the circularly polarized intensity.
Note that an individual photon is always polarized according to one of the six ways represented here.
Credits to Dan Moulton.}
\label{fig:Stokes Parameters}
\end{center}
\end{figure}

The very first Stokes parameter, $I$, gives the total intensity of the light beam.
The other three parameters give the polarized intensities in a certain direction, whose drawings can be found in \vref{fig:Stokes Parameters}.
Combining the intensities of the three different types of polarizations together gives one the so-called \textit{degree of polarization} $p=\sqrt{Q^2+U^2+V^2}/I$.
The degree of polarization gives the fraction of the total intensity that is purely polarized, ie. the intensity where one specific polarization dominates.
The amount of the total polarized intensity itself is usually referred to as the \textit{polarized intensity} $PI=p*I=\sqrt{Q^2+U^2+V^2}$.
The remaining fraction of the total intensity is called \textit{unpolarized}; all polarizations exist in equal amounts.

\begin{figure}[htb!]
\begin{center}
	\includegraphics[width=0.75\textwidth]{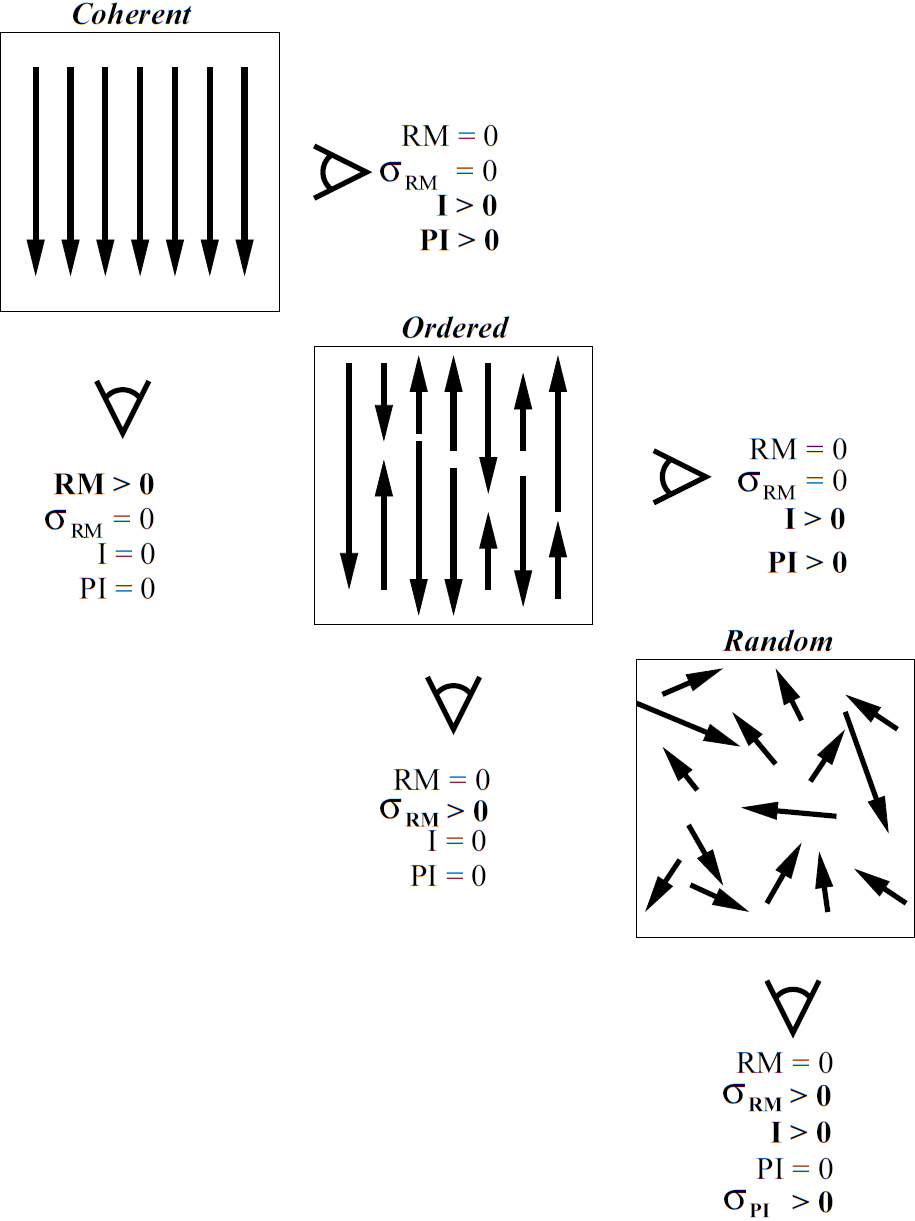}
\caption{Schematic drawing of the three different magnetic field components.
Note that the regular field is called \textit{coherent}, the anisotropic random field is called \textit{ordered} and the isotropic random field is called \textit{random}.
Also note that these three components have been discussed in a different order in the text.
The observables for these magnetic fields are the total intensity $I$, polarized intensity $PI$ and rotation measure RM.
The variance of the rotation measure $\sigma_{\mathrm{RM}}$ is also shown, since it can make a difference.
Reproduced from \textcite{txt:Jaffe}.}
\label{fig:Components Magnetic Fields Combined}
\end{center}
\end{figure}

Following the description of synchrotron radiation given in \citet{txt:Zweibel,txt:Beck}, the synchrotron radiation intensity is given by:
\begin{eqnarray}
\label{eq:Sync}
I_s&\propto&N\left(E\right)B_{\perp}^{x},
\end{eqnarray}
with $N(E)$ the density of relativistic electrons in the relevant energy range $E$ and $x$ depends on the energy spectrum of these electrons, typically $x\approx 1.8$.
However, the random magnetic field depolarizes the synchrotron radiation perpendicular to $B_{\perp}$.
The strength of this depolarization depends on the randomness of the field, which can be written as $B_{\perp, r}^2/B_{\perp, t}^2$ with $B_{\perp, r}$ the regular $B_{\perp}$ and $B_{\perp, t}$ the total $B_{\perp}$.
This means that the larger the fraction of the total magnetic field that is random, the lower the fraction of the light beam that is still polarized.
If no random magnetic field is present, all photons are polarized in the same direction.
Because of this, synchrotron radiation is excellent to use for studying random fields (see \vref{fig:Components Magnetic Fields Combined}).

Combining these two properties together allows one to calculate the strength of the magnetic field perpendicular to the line-of-sight $B_{\perp}$ (using $I$) and the fraction of the total magnetic field that is created by the regular field $B_{\perp, r}^2/B_{\perp, t}^2$ (using $PI$).
Since synchrotron radiation is always linearly polarized (circularly polarized has never been observed), Stokes parameter $V$ is not important for synchrotron radiation.

All the information in \ref{subsubsec:Faraday Rotation} and \ref{subsubsec:Synchrotron} can be summarized in a figure.
\vref{fig:Components Magnetic Fields Combined} shows the differences between the measured total intensity $I$, polarized intensity $PI$ and rotation measure RM for different lines-of-sight towards the three different magnetic field components discussed in \ref{subsubsec:Large-scale and Small-scale Fields}.
RMs measure the average strength of the magnetic field parallel to the line-of-sight, making them capable of distinguishing between regular and random field components, but unable to distinguish between the anisotropic and isotropic random field components.
Synchrotron radiation on the other hand, is perfectly capable of distinguishing between the anisotropic and isotropic random field due to isotropic random fields only having unpolarized light.
However, investigations with synchrotron radiation cannot tell regular and anisotropic random fields apart, causing them to be classified as \textit{ordered} components.
Combination of multiple lines-of-sight and tracers can make it possible to distinguish between the three different field components, but still requires a large number of assumptions.

\subsection{Galactic Magnetic Field Models}
\label{subsec:GMF Models}
Galactic magnetic field models exist in all kinds of different forms: The first tries to explain only the regular magnetic field component of the Milky Way, the second gives a detailed description of the random magnetic field of a specific area in the sky and the third wants to explain everything at once.
As pointed out in \vref{subsubsec:Large-scale and Small-scale Fields}, the random magnetic field component of the Milky Way shows large variations in its properties on 'small' scales, while the regular magnetic field component does not.
This would not be a problem in the case of modeling if the random field component only had a small contribution to the total magnetic field.
However, as stated in \vref{subsec:GMFs}, the random magnetic field is a factor of a few stronger than the regular magnetic field and therefore more 'important' to describe correctly.

Sadly, the process behind the creation of both regular and random fields is not well understood.
In fact, as said before, the best constraints for the magnetic field components are RMs and polarized synchrotron radiation, which are both line-of-sight integrated quantities.  
In this way, many assumptions about the magnetic field structure are required and thus every GMF model uses its own description of the different components.
Because of this, GMF models can differ quite a lot from each other, as models are usually made in such a way that they can explain the data set they were made for, but not other data sets.
This improves the quality of the model for that specific data set, but lowers it for all other ones, making it incompatible for comparison with other models.
This needs to be taken into account when comparing such models with each other.
Basically, the quality of a GMF model can be related to how well it describes the different components of the magnetic field with how many parameters, especially the random component.

\subsubsection{Jansson \& Farrar 2012 GMF Model}
\label{subsubsec:JF12 Model}
One of the most commonly used GMF models and the model that will be used for testing purposes in this thesis, is the model described by \citet{txt:JF12_regular,txt:JF12_random} (referred to as 'JF12' from now on).
The description of the JF12 model is split over two papers. 
The first paper \autocite{txt:JF12_regular} continues the work done in \textcite{txt:JFWE09}, and describes the methods used in the JF12 model.
Because the isotropic random field component is much more complex than the regular and anisotropic random field components, it is not treated in this paper.

The model introduces eight different spiral arm segments for the Galactic disk, all with their own regular and random field strengths.\footnote{Found in Figure 5 in \textcite{txt:JF12_regular}}
The model also uses a toroidal halo component with an out-of-plane component that is referred to as an 'X-field'.\footnote{See Figure 6 in \textcite{txt:JF12_regular}}
The anisotropic random field component is added to the model by a simple multiplicative factor.
\textcite{txt:JF12_regular} also shows some of the model performances in their Figure 2, which will be used later on for comparisons during the testing of the pipeline.
All-in-all, it introduces a $22$-parameter description of the regular Galactic magnetic field.

\begin{figure}[htb!]
\begin{center}
	\includegraphics[width=\textwidth]{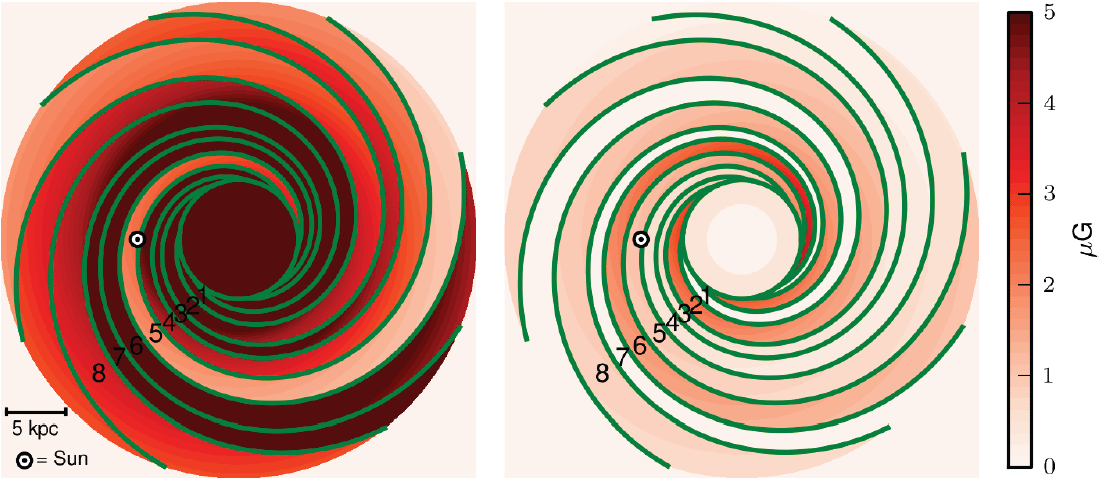}
	\caption{\textbf{Left:} Random field strength in the disk.
	\textbf{Right:} The disk component of the JF12 regular field model.
	The dot represents the location of the Sun, the green outlines the boundaries of the eight spiral arm segments and the color the magnetic field strength of either the random field (left) or regular field (right) according to the color scale as shown.
	Reproduced from \textcite{txt:JF12_random}.}
	\label{fig:JF12_random_disk}
\end{center}
\end{figure}

The second paper \autocite{txt:JF12_random} introduces a $14$-parameter model of the Galactic (isotropic) random field.
These parameters include the random components for the eight spiral arm segments, which are shown in \vref{fig:JF12_random_disk}.
The figure shows how strong and important the random field component is for the functioning of a model.
Again, a single multiplicative factor is used to account for any anisotropic random field variations.

In total, the JF12 model features $36$ parameters to describe the Galactic magnetic field: $21$ for the regular component, $13$ for the isotropic random component and $2$\footnote{One is used in the description of the regular component and one in the description of the (isotropic) random component.} for the anisotropic random component.
However, in the model that will be used for testing in this thesis (see \vref{sec:Testing}), not all of these parameters were present: Some were removed and some were manually added.
Therefore, a short description of all parameters that will be used throughout the testing phase will be given, together with the name that they have in the coded model (which will be used in all results).

\subsubsection{Hammurabi JF12}
\label{subsubsec:Hammurabi JF12}
The JF12 model used in the IMAGINE pipeline is coded into Hammurabi (see \textcite{txt:Hammurabi} and \vref{sec:Testing}).
The original coder of this model has left some parameters of JF12 out and added some manually, giving the coded JF12 model a total of $30$ parameters.

Below is a list of the parameters that are in the coded JF12 model, together with their name in both the code (used in various plots throughout this thesis) and their specific paper:
\begin{ltabulary}{|c|l|L|}
\hline
	\multicolumn{1}{|c|}{No.} &
	\multicolumn{1}{|c|}{Name} &
	\multicolumn{1}{|c|}{Description}\\
\hline
	$0-7$ & \underline{b51\_ran\_b\textit{i}} ($b_1-b_8$) & These $8$ parameters ($i=1-8$) give the field strengths of the random field component at a galactocentric radius of $5\ \mathrm{kpc}$ of the specified spiral arm segment.\\
	$8-14$ & \underline{b51\_coh\_b\textit{i}} ($b_1-b_7$) & Same as for b51\_ran\_b\textit{i}, but now for the regular field component. Note that $b_8$ is missing, since its value can be inferred from the other $7$ parameters.\\
	$15$ & \underline{b51\_z0\_spiral} ($z_0^{\mathrm{disk}})$ & Gaussian scale height of the random field component in the disk.\\
	$16$ & \underline{b51\_z0\_smooth} ($z_0$) & Gaussian scale height of the random field component in the halo.\\
	$17$ & \underline{b51\_r0\_smooth} ($r_0$) & Exponential scale length of the random field component in the halo.\\
	$18$ & \underline{b51\_b0\_smooth} ($B_0$) & Magnetic field strength of the random field component in the halo.\\
	$19$ & \underline{b51\_b0\_x} ($B_X$) & Magnetic field strength of the regular field component at the origin of the X-halo.\\
	$20$ & \underline{b51\_Xtheta} ($\theta_X^0$) & Elevation angle of the regular field component of the X-halo at the mid-plane and a galactocentric radius of $r_X^c$ (which had its best-fit value hard-coded into the model) or more.\\
	$21$ & \underline{b51\_r0\_x} ($r_X$) & Exponential scale length of the galactocentric radius of the X-halo.\\
	$22$ & \underline{b51\_h\_disk} ($h_{\mathrm{disk}}$) & Height at which the disk component transitions into the halo component.\\
	$23$ & \underline{b51\_Bn} ($B_n$) & Magnetic field strength of the regular field component in the northern region of the toroidal halo (above the Galactic plane).\\
	$24$ & \underline{b51\_Bs} ($B_s$) & Magnetic field strength of the regular field component in the southern region of the toroidal halo (below the Galactic plane).\\
	$25$ & \underline{b51\_z0\_halo} ($z_0$) & Vertical scale height of the regular field component of the toroidal halo.\\
	$26$ & \underline{b51\_b\_ring} ($b_{\mathrm{ring}}$) & Magnetic field strength of the regular field component at a galactocentric radius of $3\ \mathrm{kpc}$ to $5\ \mathrm{kpc}$.\\
	$27$ & \underline{b51\_b0\_interior} ($b_{\mathrm{int}}$) & Magnetic field strength of the random field component at a galactocentric radius of less than $5\ \mathrm{kpc}$.\\
	$28$ & \underline{b51\_reg\_b0} (Custom) & Scales the regular magnetic field strength of all different model components up or down.\\
	$29$ & \underline{b51\_shift} (Custom) & Shifts the locations where the arm segments cross the -x axis by a multiplicative amount.\\
\hline
\caption{Table containing the numbers, names and descriptions of all JF12 parameters that are used in Hammurabi.}
\label{tab:JF12}
\end{ltabulary}
All other JF12 parameters not included in this list had their best-fit values hard-coded into the model, including the two multiplicative factors for the anisotropic random field ($\gamma$ and $\beta$).
Because the last two parameters are scaling/shifting factors and represent nothing physical, they do not have an error.

\subsection{Probability Theory \& Bayesian Statistics}
\label{subsec:Probability Theory}
In order to compare GMF models with each other, one needs to be capable of describing the assumptions used in a model with a value.
One way of doing this is by using Bayesian statistics.
Here, a short overview of the basics of probability theory and Bayesian statistics is given.
A more detailed description can be found in \textcite{txt:Bayesian}.

\subsubsection{Probability Theory}
\label{subsubsec:Probability Theory}
In general, the basic algebra of probability theory is formed by the following:
\begin{eqnarray}
\label{eq:Sum Rule}
\mathrm{prob}(X|I)+\mathrm{prob}(\overline{X}|I)&=&1
\end{eqnarray}
and
\begin{eqnarray}
\label{eq:Product Rule}
\mathrm{prob}(X,Y|I)&=&\mathrm{prob}(X|Y,I)*\mathrm{prob}(Y|I),
\end{eqnarray}
with $\mathrm{prob(false)}=0$ and $\mathrm{prob(true)}=1$ defining certainty.
Here, $X$ is the proposition that a statement is true, $\overline{X}$ denotes the proposition that $X$ is false, the vertical bar '$|$' means 'given' (as in that all items to the right of this symbol are taken as being true) and the comma is read as the conjunction 'and'.
All probabilities are made conditional on $I$, to denote the relevant background information at hand.
This involves all assumptions, physical and theoretical conditions, boundaries etc. made about the proposition(s) in question.

\vref{eq:Sum Rule} is called the \textit{sum rule} and states that the probability that $X$ is true plus the probability that $X$ is false is equal to unity.
\vref{eq:Product Rule} is called the \textit{product rule}.
This states that the probability that both $X$ and $Y$ are true is equal to the probability that $X$ is true given that $Y$ is true times the probability that $Y$ is true irrespective of $X$ being true or not.

\subsubsection{Bayesian Statistics}
\label{subsubsec:Bayesian}
As said, \vrefrange{eq:Sum Rule}{eq:Product Rule} form the basic algebra of probability theory.
Many other results can be derived from these equations.
Among them are two equations\footnote{See \ref{der:Bayes' Theorem} and \ref{der:Bayes' Marginalization} for a derivation.} known as \textit{Bayes' theorem}, 
\begin{eqnarray}
\label{eq:Bayes' Theorem}
\mathrm{prob}(X|Y,I)&=&\frac{\mathrm{prob}(Y|X,I)*\mathrm{prob}(X|I)}{\mathrm{prob}(Y|I)},
\end{eqnarray}
and \textit{marginalization},
\begin{eqnarray}
\label{eq:Bayes' Marginalization}
\mathrm{prob}(X|I)&=&\int_{-\infty}^{\infty}{\mathrm{prob}(X,Y|I)dY}.
\end{eqnarray}

In Bayes' theorem, if $X$ and $Y$ are replaced by \textit{hypothesis} and \textit{data}, its usefulness becomes more clear:
\begin{eqnarray*}
\mathrm{prob}(hypothesis|data,I)&=&\frac{\mathrm{prob}(data|hypothesis,I)*\mathrm{prob}(hypothesis|I)}{\mathrm{prob}(data|I)}.
\end{eqnarray*}
All terms in Bayes' theorem have formal names and are usually written as
\begin{eqnarray}
\label{eq:Formal Bayes' Theorem}
\text{Posterior}&=&\frac{\text{Likelihood}*\text{Prior}}{\text{Evidence}},
\end{eqnarray}
which are described below.

\paragraph{Posterior}
\texttt{ }\\
The term on the left in Bayes' theorem, $\mathrm{prob}(hypothesis|data,I)$, is called the \textit{posterior} probability (or posterior probability distribution function (PDF) if one is dealing with a distribution of hypotheses).
The posterior represents the state of knowledge about the truth of the hypothesis in light of the measured data.
A more normal way of saying this, is that the posterior states how well the hypothesis can explain the measured data.

This is something that one usually wants to know: Given a certain data set, how well can my hypothesis or model explain it?
The higher the posterior probability, the better the hypothesis is at explaining the data set.
This is however something that is very hard to calculate, since the hypothesis can come in many different forms that all are incompatible with the given data set.

Therefore, something else is required to calculate this quantity that we are interested in: The likelihood.

\paragraph{Likelihood}
\texttt{ }\\
The term on the left in the nominator, $\mathrm{prob}(data|hypothesis,I)$, is commonly referred to as the \textit{likelihood}.\footnote{It is sometimes also incorrectly called the \textit{likelihood probability}. A probability implies that all terms to the right of the vertical bar are given and do not change, while the terms to the left are being tested and can be variable. Since the likelihood is calculated with a variable hypothesis, it is not a true probability.}
The likelihood differs from the posterior in that it states how well the hypothesis can predict the measured data.
The difference here is that the likelihood takes the hypothesis as being correct instead of the measured data set.
If the hypothesis is correct, then a certain data set exists that corresponds to the given hypothesis.
This data set can then be compared to the measured data set, yielding a number representing how well both data sets match with each other.

The important detail to note here, is how the hypothesis and measured data are compared with each other.
In case of the posterior, it would be required to construct a hypothesis out of the measured data.
This hypothesis is then compared to the proposed hypothesis, which only works if they both have the same form.
However, if one is capable of making such a hypothesis, one has created the perfect hypothesis already and thus the comparison makes no sense.

The likelihood however, requires one to make a data set out of the proposed hypothesis that is perfectly predicted by the hypothesis.
Since data sets usually have the same form, this data set can then be compared to the measured data set.\footnote{It is like comparing an orange with orange juice. Making an orange out of the orange juice to compare it with the other orange is impossible. Making orange juice out of the orange is not.}
The likelihood also assumes that no knowledge about the measured data is known by the hypothesis, making the likelihood test how well the hypothesis can predict the measured data.

Because of this, the power of Bayes' theorem lies in the fact that it relates the quantity of interest to a calculable quantity.
However, Bayes' theorem has two additional terms that allows one to assign a value to the hypothesis assumptions and a value to the quality of the data and the model approach.
These two terms are given by the prior and the evidence.

\paragraph{Prior}
\texttt{ }\\
The \textit{prior}, $\mathrm{prob}(hypothesis|I)$, represents the state of knowledge about the truth of the hypothesis before any probability calculations with the measured data have been made.
It includes whether or not the hypothesis satisfies all conditions set in the background information $I$.
This background information $I$ contains all conditions, assumptions, approximations, physical laws, previously acquired results and everything else that the hypothesis has to satisfy that has nothing to do with the now measured data.
In case of GMF models, a condition that one for example can put into the prior is the requirement that the model has to satisfy Maxwell's equations.
If a magnetic field of any kind does not satisfy Maxwell's equation, its model cannot exist, independent of how well this model can predict the measured data set.

\paragraph{Evidence}
\texttt{ }\\
The term in the denominator, $\mathrm{prob}(data|I)$, is called the \textit{evidence}.
The evidence is a quantity that does not depend on the hypothesis and is therefore usually omitted or treated as a normalization constant.
However, it does play a crucial role in situations like \textit{model selection}.
This is because the evidence represents the quality of both the measured data and the approach the model uses.

The importance of the quality of the measured data is trivial for model selection: When one uses data with very poor quality (large error), a model has a much easier time explaining the given data set than when it is dealing with high quality data.
This causes the evidence to go down, which leads to an increase in the posterior probability (which is logical, because low quality data sets are easily explained by models).

The evidence also describes the quality of the model approach: The higher the amount of degrees-of-freedom in a model, the lower the evidence becomes.\footnote{The prior however, also decreases when more degrees-of-freedom are introduced. In fact, it decreases faster than the evidence, meaning that the posterior probability becomes lower when the number of degrees-of-freedom increases.}
This can be explained by looking at two different models for a certain problem.
Assume that the first model uses $20$ parameters (degrees-of-freedom), while the second model uses $50$ parameters.
Now, if one assumes that both models yield the same likelihood, then it is safe to say that one would prefer the model with $20$ parameters over the model with $50$ parameters.
This is because the first model has to use more structure and needs to be more predictive than the second model, in order to predict the data equally well.
Therefore, the first model will have a higher evidence value than the second model.
Since the prior of the second model will have decreased faster than its evidence, the posterior probability of the second model will reflect the same thing (being lower than the posterior probability of the first model).

For this reason, the evidence is a very important term when one is dealing with model selection.\footnote{This is also the reason why it is given the name of \textit{evidence}, in order to capture the significance of the entity.}
Where the posterior represents how well a model can explain the measured data set independent of the actual parametrization of the model, the evidence gives the underlying information about the quality of the data and the model.

\paragraph{Bayes' marginalization}
\texttt{ }\\
Bayes' marginalization (\ref{eq:Bayes' Marginalization}) is a very powerful tool in data analysis, because it enables one to deal with \textit{nuisance parameters}: parameters which necessarily enter the analysis, but are of no intrinsic interest for the outcome.
The unwanted background signal present in many experimental measurements and instrumental parameters which are very difficult to calibrate, are examples of nuisance parameters.
Bayes' marginalization equation allows one to deal with these parameters, simplifying and lowering the amount of work needed for a Bayesian analysis.

A simple example in which one would like to use marginalization is given in Example 4 in \textcite{txt:Bayesian}:
Suppose that one has obtained a signal of interest with amplitude $A$ of a peak with known shape and position, while the background can be taken as flat and of unknown magnitude $B$.
In such a case, the PDF of the problem is given by $\mathrm{prob}(A,B|{N_k},I)$, with $A$ and $B$ the amplitudes relevant to the problem, $N_k$ the measured data points and $I$ the background information.
However, one is usually not fairly interested in the background and only in the signal $A$.
By using \ref{eq:Bayes' Marginalization}, one can marginalize over $B$ to obtain the marginal PDF for $A$:
\begin{eqnarray*}
\mathrm{prob}(A|{N_k},I)&=&\int_{0}^{\infty}{\mathrm{prob}(A,B|{N_k},I)dB}.
\end{eqnarray*}
Something that should be noted here is that the marginal PDF $\mathrm{prob}(A|{N_k},I)$ is not the same as the conditional PDF $\mathrm{prob}(A|{N_k},B,I)$.
The former PDF takes into account one's prior ignorance of the value of $B$, while the latter can be used when the magnitude of the background $B$ has already been determined in some way.
By using the marginal PDF, one does not assume a certain value for $B$, but rather just ignores it because it is not of interest.

\subsection{MCMC \& Sampling Methods}
\label{subsec:MCMC and Sampling Methods}
In statistics, one of the hardest problems to deal with is getting results from a physical or mathematical system that depends on many degrees-of-freedom in a finite amount of time.
Think for example of a system in which one wants to determine how a cosmic ray behaves as it propagates through the universe.
Such a system depends on many degrees-of-freedom that are purely random for a single cosmic ray, but are still deterministic in general.

In this case, one usually wants to use something called a \textit{Monte Carlo method}.
Monte Carlo methods are a class of algorithms that use the randomness of a system in order to solve problems, given that the problem is deterministic in principle.
By repeatedly sampling randomly over the system, the idea of Monte Carlo methods is that with enough samples, the problem can be solved by a combination of these samples.

Monte Carlo methods are usually used in three classes of problems: Numerical integration, optimization and inverse problem solving.
\paragraph{Numerical integration}
\texttt{ }\\
In systems with a large number of dimensions (degrees-of-freedom), calculating the volume of interest by usage of deterministic numerical integration algorithms can prove very difficult.
This is due to the fact that adding a dimension to the problem increases the amount of function evaluations that are required to get the wanted accuracy exponentially.
If, for example, $100$ function evaluations are required for obtaining the wanted accuracy in one dimension, then $100^{N}$ evaluations are required for $N$ dimensions.
Since physical problems can easily have $100$ or more dimensions (degrees-of-freedom), one can see that this is not going to work.

Monte Carlo methods can solve this problem quite easily as long as the function of the system is well-behaved.
By randomly sampling over the $N$ dimensions of the system, one can approximate it by taking the average of all random samples for a sufficiently large amount of samples.
By using the central limit theorem, one can see that this method has a $1/\sqrt{N}$ convergence; the error of the approximation is halved if one quadruples the amount of sampled points.
Note that this is independent of the number of dimensions.

\paragraph{Optimization}
\texttt{ }\\
Monte Carlo methods can also be used as a way of numerical optimization.
If one has a system with a problem that can have multiple outcomes, it is usually desirable to know what the best outcome is.
This can normally be calculated numerically when the system has a low number of dimensions, but becomes significantly harder when the number of dimensions increases.

This problem can be solved a lot faster by randomly sampling over all possibilities and comparing all the outcomes with each other after wards.
If the amount of samples is large enough, the best outcome is likely to be among the obtained outcomes.
Of course, the probability of finding the true best outcome depends on the number of random samples.

\paragraph{Inverse problems}
\texttt{ }\\
The third and probably most important usage of Monte Carlo methods (at least for this thesis), is its capability of solving inverse problems.
An inverse problem is the process of calculating or modeling from a set of observations the process that caused it.
For example, using RM data to calculate the magnetic field structure of the Milky Way or using the detection of a UHECR combined with knowledge about the Galactic magnetic field to backtrack the trajectory of said particle.

In general, an inverse problem is formulated in a probabilistic way, which leads to the definition of a PDF for the model space.
As discussed in \vref{subsubsec:Bayesian}, the posterior probability can be very hard to describe, which is why such a PDF is required to have.
When analyzing an inverse problem, it is usually not enough to maximize the likelihood probability, as one normally also likes to have information about the quality of the data itself (the evidence).

Since models generally tend to depend on quite a few parameters, using a marginal probability density (\ref{eq:Bayes' Marginalization}) will most likely prove very impractical or even useless.
Therefore, Monte Carlo methods can be used to randomly generate a large number of model parameters according to the PDF and test them all against the available data.

However, Monte Carlo methods do not use any information about the parameters of a problem.
Sometimes, this is not possible due to the problem having too much randomness involved.
However, if the PDF of a problem can be parametrized, which usually is the case with inverse problems, the whole process can be sped up by introducing a \textit{Markov chain} to the problem.

\subsubsection{MCMC}
\label{subsubsec:MCMC}
Doing so gives a class of algorithms that is commonly known as \textit{Markov chain Monte Carlo} (MCMC) methods.
When one is dealing with a problem that is parametrized, one knows what the functional form is of the PDF.
However, finding the maximum of this PDF can be very hard due to the number of parameters it depends on.
In order to tackle this, one needs to construct a Markov chain that eventually will have the desired PDF of a problem as its equilibrium distribution (basically converting the parametrization of the PDF to a series of numbers).
The MCMC class does exactly this, as is explained in the following.

\paragraph{Markov chain}
\texttt{ }\\
A Markov chain is a process that satisfies the \textit{Markov property}, which describes that the predictions for the future of the process are solely based on the current state of the system.
In other words, the probability of a system reaching a certain state during the next step depends only on the state it is currently in.
It can also be described as a memoryless process: the history of the system has no influence on the systems next state.

A simple example of a Markov chain can be described by a game of coin toss gambling:
Suppose that one starts with a certain amount of money and wager on a fair coin toss indefinitely or until all money is lost.
If this person has $X$ money after $n$ steps right now, one can guess that this person has either $X-1$ or $X+1$ money after the next step.
This guess is not improved in any way by having knowledge of all previous states the gambler has been in.

The process described here is a Markov chain on a countable state space that follows a random walk.
In case of calculating the model for a certain data set (an inverse problem), one has to generate many samples all undergoing their own random walk in order to find the best combination of model parameters.
However, letting all samples walk purely at random can be quite inefficient: It is perfectly possible for a Markov chain to select a state for the next step that is worse than the state it is currently in.
Although this is undesirable, it may be necessary to do if the state of a sample is currently in a local maximum.
Not allowing any state worse than the state the sample is currently in could easily result in the sample getting stuck in a local maximum of a likelihood function with many extrema.

Because one wants to get the best set of model parameters as quickly as possible, one has to find a balance between not accepting worse states and accepting worse states in order to escape local maxima.
Many different sampling methods have been created over the years in order to address this problem.
Since the IMAGINE pipeline will require a sampler that can deal with every GMF model imaginable, it is necessary to take a look at what different kinds of sampling methods exist.
The four most well-known and commonly used sampling methods are \textit{Metropolis-Hastings}, \textit{Gibbs}, \textit{Hamiltonian Monte Carlo} and \textit{Nested sampling}.

\subsubsection{Metropolis-Hastings Sampling}
\label{subsubsec:MH}
\textit{Metropolis-Hastings sampling} (MH sampling) is an MCMC method for obtaining a series of random samples from a PDF, for which direct sampling is difficult (due to many dimensions).
The algorithm is named after Metropolis, who was the main author of the first paper on this matter \autocite{txt:Metropolis}; and Hastings, who extended this idea to a more general case later on \autocite{txt:Hastings}.
The MH algorithm works by generating a sequence of sample values in such a way that, after more and more steps have been done, the distribution of values comes closer and closer to being a perfect approximation of the desired PDF.
The values of these samples are generated in an iterative way, meaning that the value of the next sample solely depends on the value of the current sample, making this sequence a Markov chain.
More specifically, the algorithm picks a random candidate for the value of the next sample based on the current sample value (random walk).
Then, by using a specified probability algorithm, this candidate is either accepted and used in the next iteration, or rejected and discarded.
The probability algorithm that determines whether or not a candidate is accepted depends on the values of the current sample and the candidate compared to the desired PDF.
Generally, if the value of the proposed candidate is better than the current value, it always gets accepted.
If it is worse, then it gets accepted with probability $\frac{P(\theta^{'})}{P(\theta)}$ with $P()$ the desired PDF, $\theta^{'}$ the proposed candidate and $\theta$ the current sample.
This way, the algorithm tends to mostly return better values, while occasionally returning worse values (which can be used to escape local maxima).

This method has a couple of disadvantages, which are shared among other MCMC methods.
The main disadvantage is that a large number of iteration steps is required before the Markov chain starts to approximate the desired PDF in an acceptable manner.
Especially if the Markov chain starts of somewhere in a minimum, it can take quite a while before it finally starts converging to the desired PDF.
The other disadvantage the MH algorithm has, is that all of its samples are correlated to each other, increasing the amount of samples required before a PDF can be approximated in an uncorrelated way.
This could be fixed by increasing the jumping width of the random walk, but this also increases the rejection rate.

Note that the above discussed method applies to one-dimensional parameter spaces.
It can easily be extended to multi-dimensional parameter spaces, in which the MH algorithm will try to choose a new multi-dimensional sample point.
However, when the number of dimensions is very high, the MH algorithm can break down.
This is due to the fact that it uses the same jumping width for every dimension, while individual dimensions generally behave themselves differently from one another.
This causes a high rejection rate and thus a slower moving Markov chain.

A solution to this problem is by only sampling one dimension at a time, not all at once.
This is known as \textit{Gibbs sampling}.
MH sampling is probably the most efficient way of sampling if the number of dimensions is low.
When the number gets higher, Gibbs sampling becomes more efficient.

\subsubsection{Gibbs Sampling}
\label{subsubsec:Gibbs Sampling}
\textit{Gibbs sampling} is, simply put, a special case of MH sampling.
Instead of sampling over all dimensions at once with the exact same jumping width (as if all dimensions depend on each other), Gibbs sampling only samples over a single dimension (or group of dependent dimensions) at once.
The individual sampling step is then in turn performed by MH sampling or something more sophisticated.
This causes less rejections to occur, since only dimensions that depend on each other, are sampled over at the same time.

To put this a bit more in perspective:
If one has a system with $N$ dimensions, of which $3$ dimensions depend on each other, then MH sampling would try to sample over $N$ dimensions at once with the same jumping width.
Gibbs sampling however, tries to use MH sampling over either these $3$ dimensions or over the remaining $N-3$ dimensions individually.
The big difference here is that independent dimensions are sampled over separately, causing random walk iterations to be accepted a lot faster and allowing the jumping widths for every dimension to be different.
It also allows for the sampler to take care of dimensions that depend on each other, by sampling over them together.
Note that this requires one to know which dimensions depend on each other, which is not always the case.

Note that Gibbs sampling is only more efficient than MH sampling if one is dealing with multiple independent dimensions that all behave in a different way.
If a system has many dependent dimensions or most dimensions behave in the same way, MH sampling will be faster than Gibbs sampling (due to the MH algorithm sampling over all dimensions at once and Gibbs doing this individually).

However, Gibbs sampling can still be inefficient.
For example, every sample point in a dimension can either have its value increased or decreased.
So, if one has a system with $N$ dimensions again, then there are basically $2^N$ possible directions a full random walk iteration step can go to (note that MH only has $2$ possible directions; either increase or decrease all values of all dimensions).
A Gibbs sampler will explore most of these $2^N$ directions, while it is not very likely that all of these will yield a better probability in the Markov chain.
In fact, only a few of these directions will be accepted.

Therefore, it would be useful if a sampling method existed that can give a prediction on the direction every dimension has to go in order to obtain a better probability.
This is given by a sampling method called \textit{Hamiltonian Monte Carlo sampling} (HMC sampling).
Gibbs sampling is very efficient when dealing with multiple, independent dimensions that all are well behaved and simple (no high amounts of extrema).
HMC sampling becomes more efficient when this is not the case.

\subsubsection{Hamiltonian Monte Carlo Sampling}
\label{subsubsec:HMC}
\textit{Hamiltonian Monte Carlo sampling} (also known as \textit{Hybrid Monte Carlo sampling}) is a unique MCMC algorithm that is capable of generating a so-called vector field with the most probable direction at every point in parameter space.
Instead of trying random directions for random jumps in parameter space, one can follow the direction that is assigned to every point for a small distance.
This will then move the sample point to a new point in parameter space, which has a new direction to follow.
Continuing this trend of following the directions of the vector field allows for fast movement on the PDF.

The big question that still remains here however, is how one is going to construct a vector field that is aligned with the most probable direction in parameter space, by only using information that can be gained from the PDF.
One piece of information that the previously discussed MCMC methods do not use, is the differential structure of the desired PDF, which can be obtained through the gradient of the desired probability density function.
This gradient will define a vector field in parameter space that is sensitive to the structure of the desired PDF.

The problem with this sensitivity is that the gradient is never aligned with the true desired PDF, but rather with its density.
This means that the gradient captures the probabilistic structure that depends on the parametrization, but not the structure that is invariant under it.
Therefore, extra information and geometric constraints are required to use this gradient correctly.
According to \textcite{txt:Betancourt}, the differential geometry that is required to correct these density gradients also happens to be the mathematics that describes classical physics.
Therefore, one can say that for every probabilistic system there is a mathematically equivalent physical system that is easier to describe.

By using this idea, one can expand the $N$-dimensional parameter space to a $2N$-dimensional phase space by introducing additional momentum parameters,
\begin{eqnarray}
\theta_n &\rightarrow& \left(\theta_n, p_n\right),
\end{eqnarray}
with $\theta$ representing the unknown parameter(s) and $p$ the additional momentum parameter(s).
Now that the parameter space is expanded to phase space, one can lift the desired PDF onto a PDF in phase space, which is commonly known as the \textit{canonical PDF}.
Doing so with the choice of a conditional PDF over the additional momentum, gives
\begin{eqnarray}
\mathrm{prob}(\theta, p)&=&\mathrm{prob}(p|\theta)*\mathrm{prob}(\theta),
\end{eqnarray}
which also ensures that the momentum can be removed by simply marginalizing over it.

Because of the duality of the parameters and the momentum, the corresponding probability density functions also transform oppositely to each other.
In fact, the canonical probability density $\mathrm{prob}(\theta, p)$ does not depend on the parametrization at all, which means it can be written in terms of an invariant \textit{Hamiltonian} function $H(\theta, p)$,
\begin{eqnarray}
\mathrm{prob}(\theta, p)&=&e^{-H(\theta, p)}.
\end{eqnarray}
Because $H(\theta, p)$ is independent of the parametrization, it captures the probabilistic structure of the phase space PDF that is invariant (which was not possible in parameter space).
Now that the geometry of the system has been captured with a Hamiltonian, one can simply use \textit{Hamilton's equations} in order to generate a vector field that can predict which direction is more favorable.
A more detailed description of the mechanics of HMC sampling can be found in \textcite{txt:Betancourt}.

HMC sampling works very well when one is dealing with a large number of dimensions, especially if these are all independent of each other and behave very differently.
Because HMC sampling can give a prediction on which direction a sample point in parameter space has to go, it can heavily reduce the amount of time taken to perform a successful Markov chain iteration step.
However, as with all MCMC methods discussed up till now, HMC fails when dealing with PDFs that are not well-behaved.
MH and Gibbs fail because these methods cannot recognize a discontinuity, while HMC fails because the derivative of a discontinuity is not defined and thus breaks the algorithm.
Another thing is that, like the others, HMC starts at a single point in parameter space and generates a Markov chain from there.
It is still quite susceptible to local maxima.

A solution to this problem would be the usage of \textit{nested sampling}.
Nested sampling is much slower than all MCMC methods discussed, especially HMC, but in return it is the only MCMC method that practically cannot get stuck. 

\subsubsection{Nested Sampling}
\label{subsubsec:Nested Sampling}
\textit{Nested sampling} is a unique MCMC method that is, according to \textcite{txt:Skilling}, capable of directly estimating the relation between the likelihood function and the prior mass.
It is also unique in the fact that nested sampling lets one immediately obtain the evidence (see \vref{subsubsec:Bayesian}) by summation.
Nested sampling differs from other sampling methods in that it uses the evidence as its primary result, unlike the posterior probability like most sampling methods do.
This allows the sampler to be capable of comparing different model assumptions (and thus entirely different models) through the ratios of evidence values (which are known as Bayes factors).
Presenting the evidence value lets the results of different model optimizations be future-proof, in that future models can be compared with the current one without having to re-do the current calculations every time a new model pops up. 

Nested sampling also works in a different way than the other MCMC methods described thus far.
The other MCMC methods start with a single sample somewhere in parameter space and create a Markov chain from that point.
This has the weakness that if this sample point starts of somewhere with a low probability density or is close to a local maxima, it can have a hard time getting out.
The other weakness of only starting with a single sample point is that it cannot deal with discontinuous likelihood functions, something that can be encountered in model optimizations (see \vref{sec:Testing}).
Of course, all these MCMC methods can start of with multiple sample points in parameter space, but each of these points will still create their own Markov chain, which still has these weaknesses.

Nested sampling does not suffer from this, because it does not try to sample over parameter space, but over likelihood space.
A nested sampler creates a large amount of samples to start with (depending on the number of dimensions) and creates sort-of mini-Markov chains with them.
Instead of generating a Markov chain for every sample point in parameter space (like other MCMC methods), a nested sampler tries to improve the sample point with the lowest likelihood at every iteration step.

This procedure can be summarized like the following:
Assume that one starts with $N$ sample points in parameter space, called $\theta_1, ..., \theta_N$.
Every $\theta$ has its own likelihood value, called $L(\theta_1), ..., L(\theta_N)$.
At every iteration step $i$, there is one $\theta$ that gives the lowest likelihood, which is called $L_i$.
Now, replace the $\theta$ with likelihood $L_i$ with a random other $\theta$ in the collection.
Perform a random walk Markov chain on the replaced $\theta$ in order to create a new sample point, keeping in mind that $L(\theta_{\mathrm{new}})>L_i$.
Make this Markov chain $M$ iteration steps long and start over again after a new sample point has been created.

Looking at the procedure described above, one can easily see that this differs quite a lot from other MCMC methods.
The most notable thing is that a nested sampler replaces sample points by other points that are not near the replaced sample point, like other MCMC methods do.
Instead, it replaces the sample point with the lowest likelihood by a random other sample point.
This ensures that the likelihood of the new sample point is higher than the likelihood of the replaced one.
It does however not ensure that a new sample point is created, since a sample point was simply copied.
That is what the random walk Markov chain is for: Creating a new sample point to replace the older one.
This Markov chain is usually created by using Gibbs sampling with an adaptive jumping width.
The adaptive jumping width increases the probability of getting away from the starting point of the Markov chain faster, by increasing it if the acceptance ratio increases and decreasing it when the ratio decreases.

This method has a couple of advantages and disadvantages over other MCMC methods.
Something one can notice in the description given above, is that a nested sampler always replaces the sample point with lowest likelihood with a better one.
This means that, unlike other MCMC methods, sample points with lower likelihoods are never accepted.
Most MCMC methods are made in such a way that they can approximate the desired PDF, something a nested sampler simply cannot do.
A nested sampler exists purely to find the maximum likelihood in a PDF and is therefore often called the sampling method specialized for Bayesian statistics (since one only cares about the highest probability in Bayesian).

An other disadvantage of nested sampling, is that it is very slow in comparison to HMC.
Especially with a high number of dimensions, nested sampling requires a couple of hundreds or thousands samples before it can even start.
Not enough starting samples, and nested sampling can occasionally remove good sample points on a weird likelihood function.
Too many and it can take a nested sampler a long time before it reaches the maximum.

These disadvantages however, usually do not weigh against the big advantage nested sampling has over other MCMC methods, if used correctly: It cannot get stuck.
If an MH or HMC sampler meets a discontinuous function or a function with many extrema, it can easily get stuck (especially in discontinuous functions).
A nested sampler however cannot, due to the fact that it does not create a Markov chain of random walks from a single sample point, but creates many small Markov chains from sample points all over parameter space.
If a sample point is somewhere where an MH or HMC sampler gets stuck, a nested sampler will forget about that point and replace it later when it becomes the sample point with the lowest likelihood.
This makes nested sampling the superior choice when one is dealing with PDFs with complex behavior.

In conclusion, the four MCMC methods can be summarized in the following way:
For simple problems with a low number of dimensions, MH sampling is the best choice due to its simplicity and basically no preparation time.
When the number of dimensions becomes higher and multiple dimensions become independent, Gibbs sampling becomes the better choice due to its individually handling of dimensions.
If the number of dimensions becomes even higher than that and the PDF is well-behaved, then HMC sampling is the superior sampler.
And, finally, if the desired PDF is not a well-behaved function, then nested sampling is the best MCMC method of all.

Since the IMAGINE pipeline will have to deal with GMF models, which usually have a large number of parameters (dimensions), the question here is whether or not it can be expected that the PDF of a GMF model is well-behaved.
If it is, HMC sampling is always superior over nested sampling.
If not, nested sampling is the only MCMC method left that can deal with the problem.

\newpage
\section{IMAGINE Pipeline}
\label{sec:IMAGINE Pipeline}
\subsection{Introduction}
\label{subsec:IMAGINE Introduction}
Using the Bayesian probability theory discussed in \vref{subsec:Probability Theory} as a basis, the \textit{Interstellar MAGnetic field INference Engine} (IMAGINE) pipeline is being developed.
The IMAGINE pipeline is created in order to be capable of optimizing and testing various GMF models, making Bayesian statistics a core ingredient.
Usually, models cannot be compared to each other due to them using different data sets.
A model (of any kind) is generally made such that it is capable of explaining the data set is was made for, but is incapable of explaining other data sets (which is basically an assumption).
Using Bayesian statistics, a quantity can be assigned to these different data sets (the evidence), allowing models to be compared with each other and also making it possible for every model to be tested against every data set.

\begin{figure}[htb!]
\begin{center}
	\includegraphics[width=\textwidth]{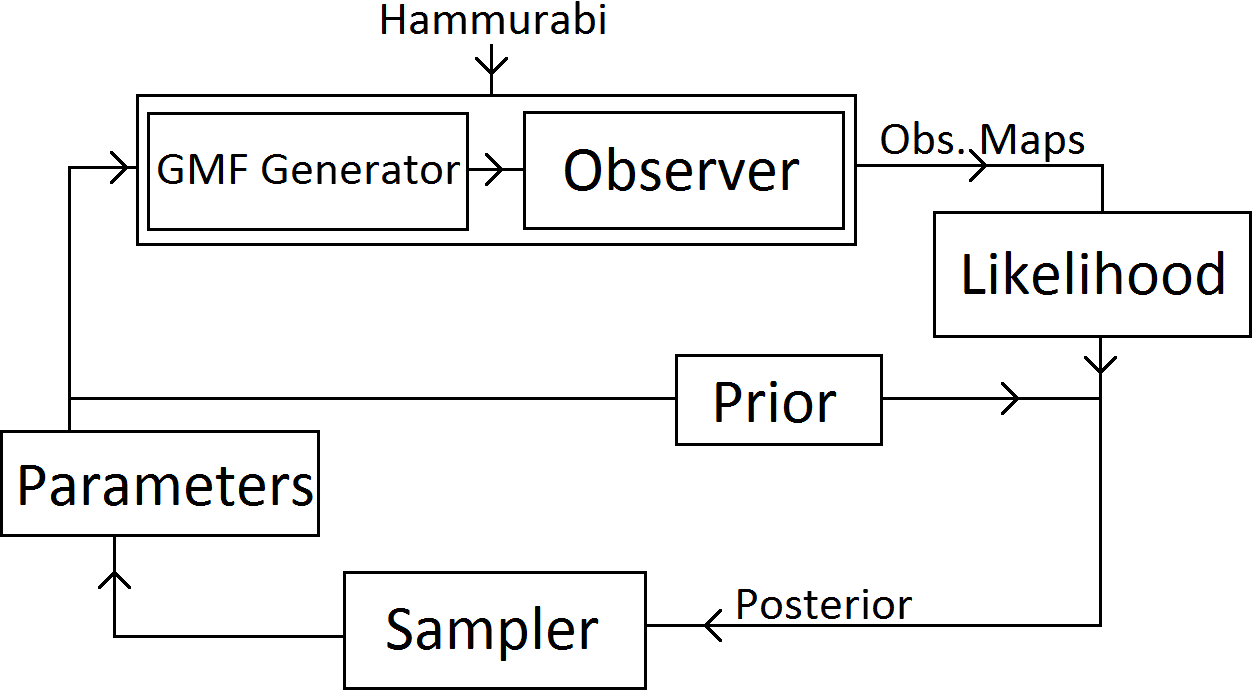}
	\caption{A schematic overview of the IMAGINE pipeline.}
	\label{fig:IMAGINE pipeline}
\end{center}
\end{figure}

A schematic overview of the IMAGINE pipeline can be found in \vref{fig:IMAGINE pipeline}.
The pipeline is made up by a set of modules.
The idea of the pipeline is that these modules all work independently of each other; one can swap out a module for an other and the pipeline still works perfectly fine.
This adds a lot of versatility to the pipeline, since researchers can use their own algorithms in the pipeline.
IMAGINE itself basically only provides the connection between the modules and creates a framework for Bayesian statistics.

However, this versatility comes at the price that all modules have to work completely independently from each other, which makes them harder to create.
Below is a short description of all individual modules of IMAGINE.\footnote{These descriptions tell how IMAGINE should work on paper before the work of this thesis started. Not everything was/is already implemented in the pipeline. Descriptions are also subjected to changes in the future.}

\paragraph{Sampler}
\texttt{ }\\
The pipeline starts at the \textit{sampler} module.
Here, the pipeline receives the parameter set of the GMF model that requires optimization.
The sampler module contains one or several samplers based on the MCMC methods discussed in \vref{subsec:MCMC and Sampling Methods}.
Depending on the number and complexity of the parameters of this GMF model, the pipeline can pick the MCMC method it thinks is required to do the optimization as quickly as possible.

The sampler module is probably the most important module in the IMAGINE pipeline.
If implemented incorrectly, the whole pipeline will not work.
For that specific reason, the work done in this thesis is mainly about checking what exactly a sampler must be capable of doing and how it must be implemented in the pipeline.

\paragraph{Prior}
\texttt{ }\\
The sampler sends the values of all parameters to a module that can calculate the \textit{prior}.
Although quite simple, this module is really important since it represents the state of knowledge of the model before any comparisons with data are made.
This state of knowledge is useful, because it can take care of the validity of the model compared to the conditions one sets for it.
In case of GMF models, one minimal condition every magnetic field has to fulfill, is satisfying Maxwell's equations.
If a magnetic field does not satisfy Maxwell's equations, the GMF model with the current parameter values cannot be correct, independent of the data it can generate.

\paragraph{GMF Generator}
\texttt{ }\\
The sampler also sends the values of all parameters to the \textit{GMF generator} module.
This module uses the received values for all the parameters in order to create a representation of the GMF that the model describes.
The GMF generator is capable of interpreting the way a GMF model works and generate a map with the magnetic field values of the galaxy from it.
The generation of GMF maps is currently handled by the Hammurabi code \autocite{txt:Hammurabi}.

\paragraph{Observer/Observable Generator}
\texttt{ }\\
The generated GMF is then send to the \textit{observer} (or \textit{observable generator}) module.
This module takes a GMF map and generates artificial data maps with it called \textit{observable maps}.
These observable maps contain data from different types of tracers that one would detect, if the given GMF map is correct.
For example, for a given GMF map, one would expect to observe certain values for the RM or the synchrotron Stokes IQU-parameters for every magnetic field data point in the GMF map.
This module generates these RM and IQU-maps by using the Hammurabi code from \textcite{txt:Hammurabi}.

\paragraph{Likelihood}
\texttt{ }\\
The artificially generated observable maps can then be used in the \textit{likelihood} module.
Following the knowledge of Bayesian statistics in \vref{subsec:Probability Theory}, this is required in order to calculate the likelihood of the model.
This module is most likely the most important module after the sampler, since many different mathematical ways can be used to calculate how well the observable maps compare to the real data.
Some ways are also explored and tested in the work done in this thesis.
Not surprisingly, the likelihood module only works correctly if enough real data is available.

\paragraph{Posterior}
\texttt{ }\\
And, finally, the calculated values for the prior and the likelihood can be sent to the \textit{posterior} module.
This module combines the prior, the likelihood and the evidence to obtain a value for the posterior probability.
Note that there is no \textit{evidence} module in \ref{fig:IMAGINE pipeline} yet, because it was not deemed important to have an individual module for the evidence at the start of the work of this thesis.

After a value for the posterior probability has been found, it is fed back into the sampler module.
The sampler module will then go to the next iteration step in its Markov chain.
This loop will continue until either manually stopped or the sampler itself stops according to some given requirements.

The power of the IMAGINE pipeline, is that it is a basic pipeline made up by modules.
Each module performs a specific task, and the pipeline combines all these tasks together to make it work.
This means that all modules in the pipeline are independent of each other, and can perfectly work without any of the other modules.
Therefore, any module can be interchanged with a different one, as long as it performs the same task.
This gives the pipeline more versatility, because it simply connects all modules of choice with each other.
If one is for example not satisfied with the sampler the pipeline currently uses, a different sampler can be used as long as it is made compatible with the inputs and outputs of the pipeline. 

\subsection{Development}
\label{subsec:IMAGINE Development}
As described in the previous section, the IMAGINE pipeline consists out of individual, independent modules that work together in order to create the functionality of the pipeline.
However, all these different modules need to be built, implemented, tested and optimized.
Since the IMAGINE pipeline project already existed for multiple years before the research in this thesis was done, some components of the pipeline were already built (be it in a pre-release state).\footnote{For example, the overall structure that connects all modules together already existed.}
Since no work was done on these components throughout this thesis research, nothing will be reported about them.

In the current state of IMAGINE, all modules already exist in the pipeline.
However, the sampler module only has a rough implementation of an HMC sampler and the likelihood module can only use $\chi^2$-optimizations\footnote{A $\chi^2$-optimization is the process of minimizing the sum of the squared differences between the data points in both data sets.} at this point.
As stated before, these two modules are arguably the most important modules of the IMAGINE pipeline and therefore need extensive testing and a good implementation.

According to \vref{subsubsec:Nested Sampling}, an HMC sampler is very efficient when dealing with large number of dimensions/parameters in models, but only works if all parameters are well-behaved.
If not, a nested sampler is required for the pipeline, although much slower than an HMC sampler.
Also, one needs to investigate if a simple $\chi^2$-optimization is sufficient enough for model optimization.
Since $\chi^2$-optimization basically gives a predetermined weight to every data point, it could potentially fail to distinguish important data points from less important points.

In \ref{sec:Testing}, I report my findings on the research I have done on these two modules: The likelihood and the sampler module.

\newpage
\section{Testing}
\label{sec:Testing}
The main goal of the research done in this thesis, was to check if the IMAGINE pipeline was ready to start optimizing GMF models and what kind of sampler would be required in order to do such a task.
In order to check if the pipeline is ready, a GMF model is of course required to already be implemented in the pipeline.
A perfect candidate for this was the JF12 model (see \vref{subsubsec:JF12 Model}), since this is a commonly known and widely used model.
This model was also the most sophisticated model present in the Hammurabi code, which was already implemented in the IMAGINE pipeline at that time. 

One of the most important parts of the pipeline that was still missing at the start of the performed tests, was the sampler.
It was not really known yet what kind of sampler was required in order to successfully, but also time efficiently optimize a GMF model.
A couple of tests were already performed with an HMC sampler (see \vref{subsubsec:HMC}), but these were unsuccessful.

This section contains information about the tests that have been done with the IMAGINE pipeline, the results that were obtained and the problems that were discovered.
\ref{subsec:Testing Details} shows some details on what exactly was being tested with what kind of result, what assumptions were made during the tests and some discussion.
The remainder of this section is divided up into three parts.
\ref{subsec:Likelihood Plots Part 1} to \ref{subsec:Discussion Part 1} show the tests I have done on the JF12 implementation in Hammurabi called model $51$.
After that, I explore various other aspects about the JF12 model and Hammurabi that can improve the sampler module in \ref{subsec:Model 7 and Model 51} to \ref{subsec:NE2001 or Gaensler?}.
\ref{subsec:Likelihood Plots Part 2} to \ref{subsec:Discussion Part 2} take the acquired knowledge of the previous tests into account and show the tests I have done on the JF12 implementation in Hammurabi called model $7$.
Something that should be stated here (and can also be found in the respective discussions), is that most of the problems that are shown in this section are no longer present in the pipeline.
If this is the case, then that simply means that they have already been solved or that they were not a problem after all.

\subsection{Testing Details}
\label{subsec:Testing Details}
\begin{figure}[htb!]
\begin{center}
	\includegraphics[width=0.5\textwidth]{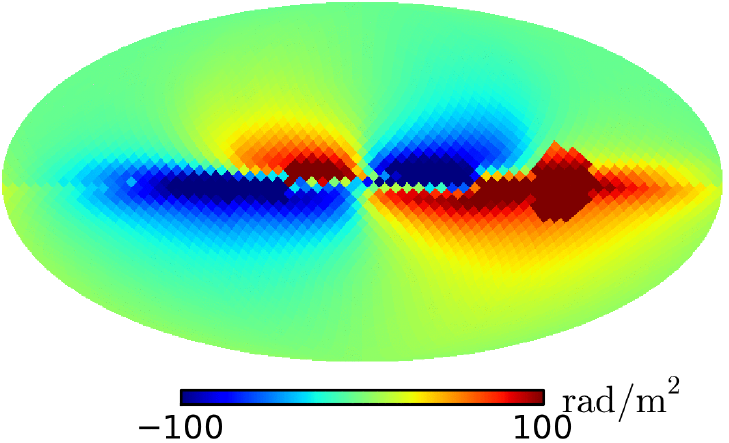}
	\caption{RM map obtained by using the JF12 model. Reproduced from \textcite{txt:JF12_regular}.}
	\label{fig:JF12 RM}
\end{center}
\end{figure}

Since previously was stated that the JF12 model will be used in order to test the IMAGINE pipeline, some JF12 model data or results is required.
The first JF12 paper \autocite{txt:JF12_regular} has a map showing the RM-values that should be obtained by using their model with Hammurabi.
This RM map can be found in \vref{fig:JF12 RM}.

Since the same JF12 model is implemented in Hammurabi in IMAGINE, the pipeline should be capable of recreating the JF12 RM map if the whole pipeline is working as intended.
Therefore, a simple test was introduced to utilize this in order to test the pipeline for the right sampling method.

\paragraph{Simple test}
\texttt{ }\\
The JF12 papers contain a list of values for the parameters in the model.
By choosing these values as the default values for all JF12 parameters and by using Hammurabi, mock RM data can be created of the JF12 model.
This mock data can then be used as the real/comparison data in the IMAGINE pipeline.
This has an advantage over using real data, because real data is unpredictable.
It also acts as a consistency check, because it is already known that Hammurabi can create the mock data and should thus also be capable of creating it again.

Now, by using this mock data, one can check what the behavior of the posterior PDFs of the JF12 model are.
According to \vref{subsec:MCMC and Sampling Methods}, an HMC sampler is very time-efficient when one is dealing with a large number of dimensions.
It can however not deal with discontinuous functions.
If such a thing would occur, then the only remaining sampler that can be used in the pipeline, is a nested sampler.

These tests were specifically done to check which of the two samplers is better for the pipeline.
However, since checking the behavior of all posterior PDFs for all $30$ parameters is very time-consuming, a couple of assumptions have been made:
\begin{itemize}
	\item \underline{Set prior to unity.} This basically means that everything is known about the model and simplifies the calculations;
	\item \underline{Set evidence to unity.} This then also means that everything is known about the data;
	\item \underline{Likelihood is a $\chi^2$-optimization.} This is one of the simplest ways of calculating how well a fit/model compares to the data;
	\item \underline{Parameter values are considered in a $3\sigma$-range.} $3\sigma$ takes $99.7\%$ of all possible values into account, which is more than enough for simple testing.
\end{itemize}
The first and second point together mean that the posterior in Bayes' theorem (\ref{eq:Bayes' Theorem}) is now equal to the likelihood.
For this reason, the term 'likelihood function' or 'likelihood plot' will be used to address the special posterior PDFs.
This also simplifies the optimization calculations quite a lot, but can cause numerical artifacts (see \vref{sec:Discussion and Conclusion}).

Basically what one wants to look at during the testing of the likelihood functions, is whether or not they are continuous.
Continuous functions are fairly easy to optimize with almost any sampling method.
However, discontinuous functions can be really hard or maybe even impossible to optimize.
Whenever a function shows discontinuous behavior, it is important to check what the reason for this is: a physical reason or a numerical reason?
If discontinuity is caused by a physical effect, then this effect can be described in a certain (physical) way and thus can be given to the sampler.
The sampler can then take this as an extra condition and 'knows' to expect discontinuity at certain positions in the likelihood function.
If the discontinuity is caused by a numerical effect however, then a sampler cannot effectively deal with it, since there is no way of telling the sampler how the discontinuity will look like.

So, all $30$ parameters of the JF12 model need to be checked for this.
To do this in an efficient way, something called a \textit{carrier mapper} has been used to check all possible values for all parameters.

\begin{figure}[htb!]
\begin{center}
	\subfloat{\label{subfig:Carrier Mapper}\includegraphics[width=0.49\textwidth]{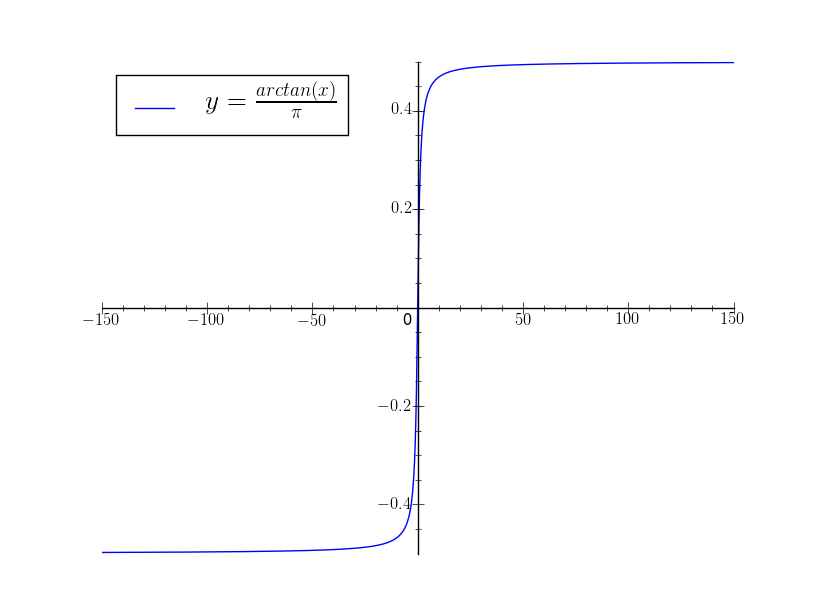}}
	\subfloat{\label{subfig:Example Likelihood Plot}\includegraphics[width=0.49\textwidth]{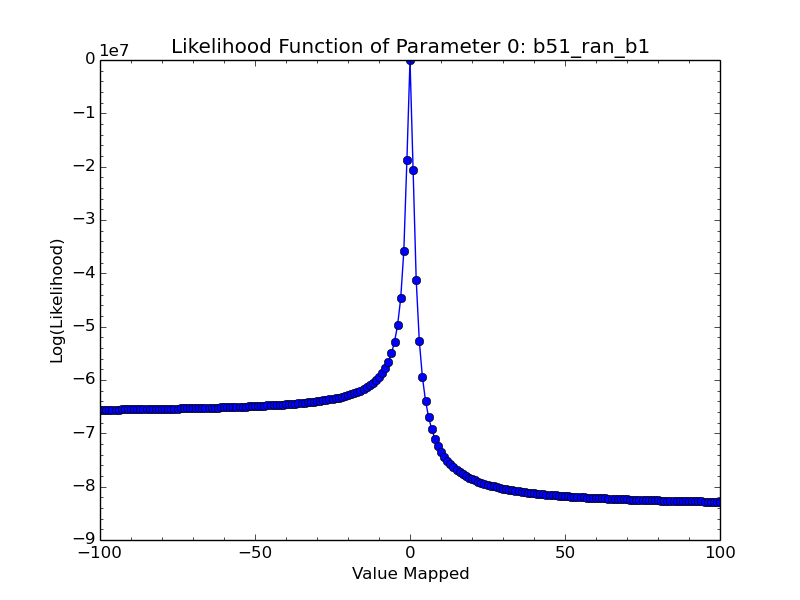}}
	\caption{\textbf{Left:} Plot showing the $\frac{\mathrm{arctan}(x)}{\pi}$-function that the carrier mapper uses.
	\textbf{Right:} An example of a likelihood plot that will be used throughout this thesis.
	The title states to which parameter this likelihood plot belongs to, the $x$-axis the value that was given to the carrier mapper and the $y$-axis the natural logarithm of the likelihood.
	In this plot, $29$ out of the $30$ JF12-parameters were set to their default value, while the remaining parameter was being varied.}
	\label{fig:Carrier Mapper}
\end{center}
\end{figure}

\paragraph{Carrier mapper}
\texttt{ }\\
As said in the previous paragraph, all parameter values are considered in their $3\sigma$-range.
However, the values inside this range are not chosen at random, but instead are chosen by the carrier mapper.
The carrier mapper is a function that basically maps all values in the $x=[-\infty, \infty]$-range to a finite $[-0.5, 0.5]$-range by usage of an $\mathrm{arctan}$-function.
This function can be seen on the left in \vref{fig:Carrier Mapper}.
By setting $-3\sigma$ to $-0.5$, $+3\sigma$ to $0.5$ and the default value to $0$, the $\frac{\mathrm{arctan}(x)}{\pi}$-function can be used to effectively scan over all possible values.
The carrier mapper takes a value $x$ in and maps this in such a way that a low amount of steps is required to still accurately cover a wide range.
This is because the carrier mapper assumes that most of the values will be close to the default value at $x=0$ and only a few values are far away.
Therefore, the carrier mapper maps most values far away from the default value, which makes it much easier to rule these points out.
This is shown by the fact that the range of $x=[-1, 1]$ already covers the $1.5\sigma$-range.

Since the carrier mapper takes in values between $-\infty$ and $\infty$, the likelihood plots (an example can be seen on the right in \ref{fig:Carrier Mapper}) show the value that was given to the carrier mapper on the $x$-axis.
In here, a value of $0$ equals the default value for that specific parameter, a negative value equals the default value minus a certain amount of sigma's and vice versa.
Since it is computationally impossible to use all values between $x=-\infty$ and $x=\infty$, a range of $x=[-100, 100]$ has been used throughout this thesis.
As shown in the plot of the carrier mapper, a range of $x=[-100, 100]$ almost completely covers the total range.

\subsection{Likelihood Plots: Part 1}
\label{subsec:Likelihood Plots Part 1}
It is important to check how the parameters of a model behave when they are assumed to be independent of each other.
In this way, if parameters are dependent on each other, they most likely will show discontinuous behavior and will be detected.
Therefore, all likelihood plots in this thesis (like \ref{subfig:Example Likelihood Plot}) show the likelihood function of a single parameter.
In these plots, a single parameter was varied between mapped values $x=[-100, 100]$ and the remaining $29$ parameters were set to their default values (which is mapped to $x=0$).
As we compare the model to mock data created with the same model, the natural logarithm of the likelihood equals $0$ (which is a likelihood of $1$) at $x=0$ (see \ref{subfig:Example Likelihood Plot}).
This should be the case in every plot in this thesis, and acts as a consistency test.
A continuous likelihood should show up as a single peak around $x=0$.

The first series of tests were performed by using only default values in Hammurabi.
These tests were meant as a starting point.
The Hammurabi code has some parameters itself as well, which were not touched during these tests unless specifically stated.

\begin{figure}[htb!]
\begin{center}
	\subfloat{\includegraphics[width=0.33\textwidth]{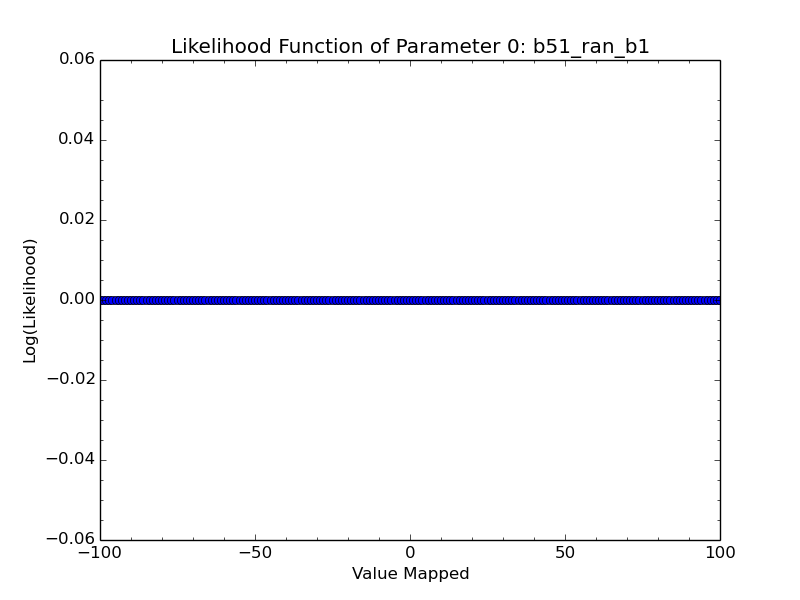}}
	\subfloat{\includegraphics[width=0.33\textwidth]{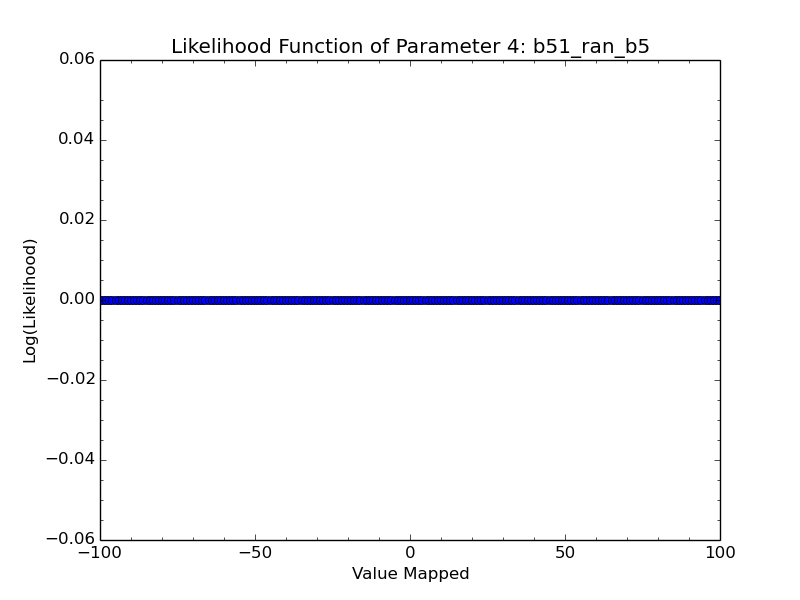}}
	\subfloat{\includegraphics[width=0.33\textwidth]{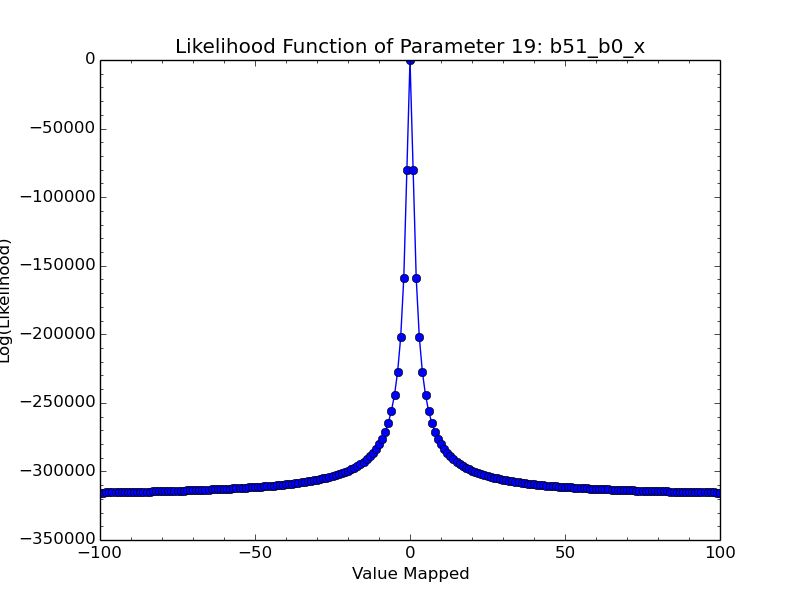}}
	\caption{Likelihoods plots of parameters $0$ and $4$ (giving the strength of the random magnetic field in two spiral arm segments) and parameter $19$ (giving the strength of the regular field of the X-halo) of the JF12 model with no random field enabled and usage of Hammurabi model $51$.}
	\label{fig:Part 1 no random}
\end{center}
\end{figure}

Likelihood plots were made of all $30$ JF12 parameters that were implemented.
The resulting plots of parameters $0$, $4$ and $19$ can be found in \vref{fig:Part 1 no random}.

\begin{figure}[htb!]
\begin{center}
	\subfloat{\includegraphics[width=0.33\textwidth]{Individual_likelihood/fig0_RM.png}}
	\subfloat{\includegraphics[width=0.33\textwidth]{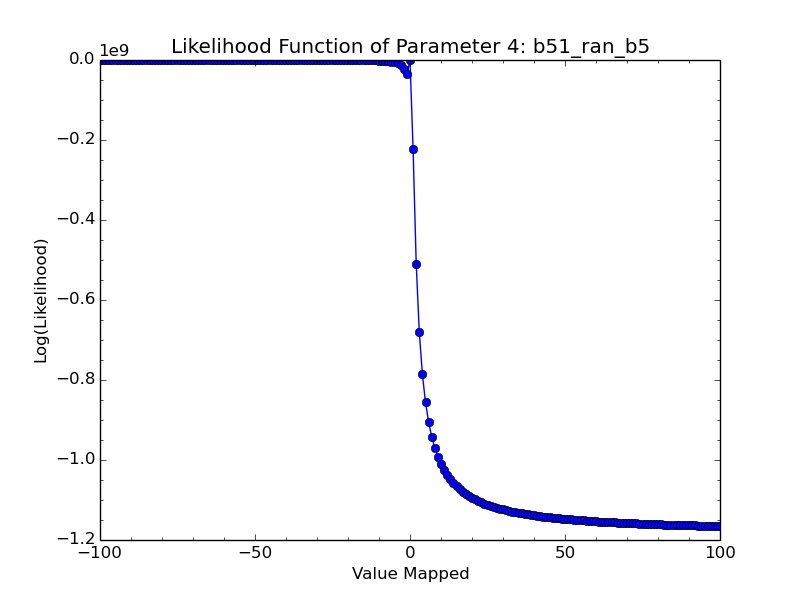}}
	\subfloat{\label{subfig:Parameter 19 do random Part 1}\includegraphics[width=0.33\textwidth]{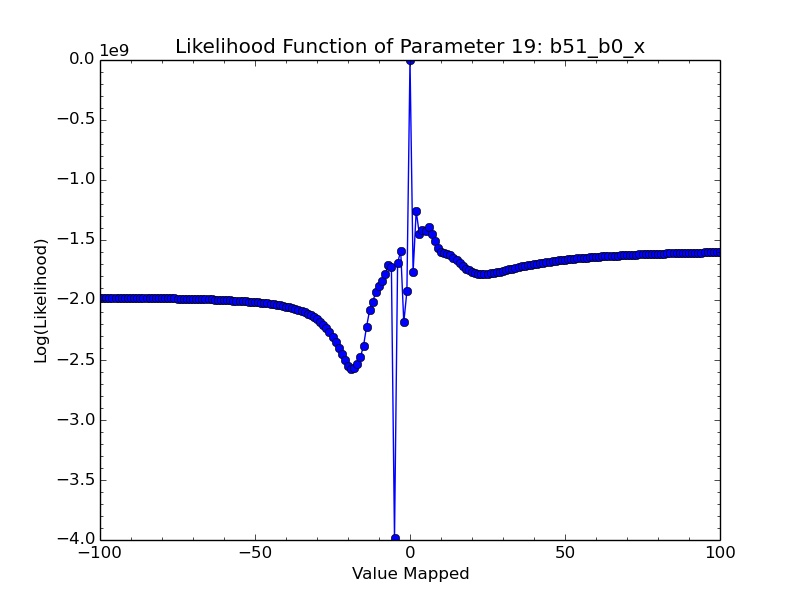}}
	\caption{Likelihoods plots of parameters $0$, $4$ and $19$ of the JF12 model with a random field enabled and usage of Hammurabi model $51$.}
	\label{fig:Part 1 do random?}
\end{center}
\end{figure}

The default Hammurabi settings have no random field enabled.
Therefore, the likelihoods denoting random field strengths in \ref{fig:Part 1 no random} are zero everywhere.
This holds for all $13$ random field parameters of JF12.
The likelihood plot of parameter $19$ show really continuous behavior, exactly what one wants from a likelihood function.
When the random field is turned on, the likelihood plots of the same parameters are as shown in \vref{fig:Part 1 do random?}.

Looking at the likelihood plot of parameter $0$ in \vref{fig:Part 1 do random?}, it behaves in a nice and continuous way now that the random field is enabled.
It also shows that the probability of the parameter being below its default value is higher than being above it.

The likelihood plot of parameter $4$ however, shows something interesting: For positive mapping values, it shows a normal fall-off.
For negative values, it only shows the start of a fall-off and then immediately goes back to almost the top of the plot.
This can be explained by looking back at how the carrier mapper works: It creates only a few points close to the default value, while making many points far away from it.
Parameter $4$ has a positive value by default, but can become negative if multiple sigma's are subtracted from the default value.
Since this also changes the direction of the magnetic field, it can impact the way the likelihood function behaves.
And thus, according to this likelihood plot, parameter $4$ has a much higher probability of having a negative value than a positive one.

Finally, the likelihood plot of parameter $19$ shows some very discontinuous behavior around the default value.
The problem here is that this discontinuity cannot be explained.
As can be found in \vref{subsubsec:JF12 Model}, parameter $19$ controls the strength of the regular magnetic field of the X-halo that the JF12 model introduced.
Since this parameter only influences the regular magnetic field strength, it should not show any changes when the random magnetic field is turned on.
This calls for a closer look.

\subsection{Likelihood Movies: Part 1}
\label{subsec:Likelihood Movies Part 1}
Every single data point in a likelihood plot corresponds to a full RM map.
Therefore, the RM maps of parameter $19$ have been studied to discover why it shows discontinuity near its default value.
\begin{figure}[htb!]
\begin{center}
	\subfloat{\includegraphics[width=0.33\textwidth]{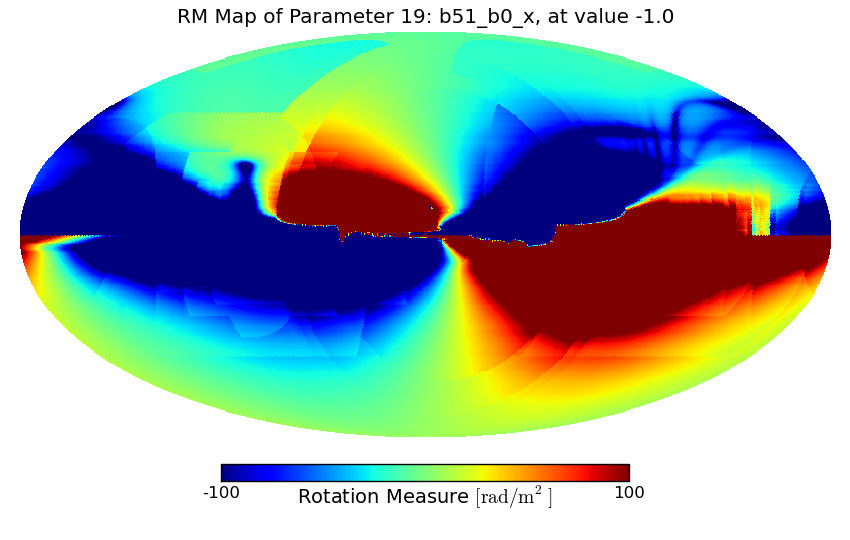}}
	\subfloat{\includegraphics[width=0.33\textwidth]{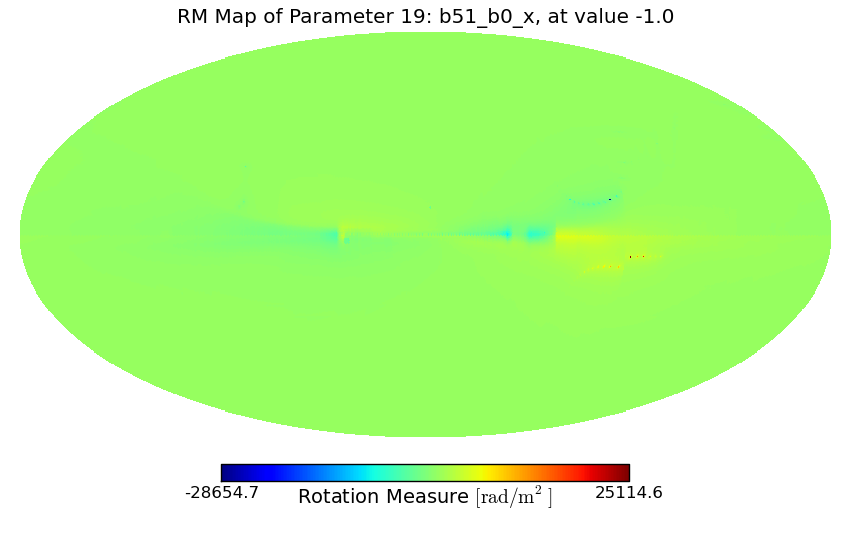}}
	\subfloat{\includegraphics[width=0.33\textwidth]{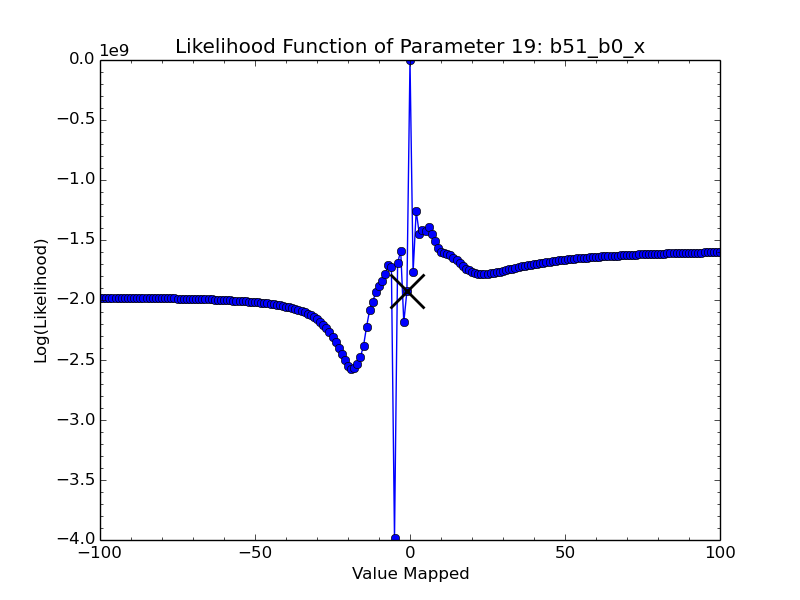}}\\
	\subfloat{\includegraphics[width=0.33\textwidth]{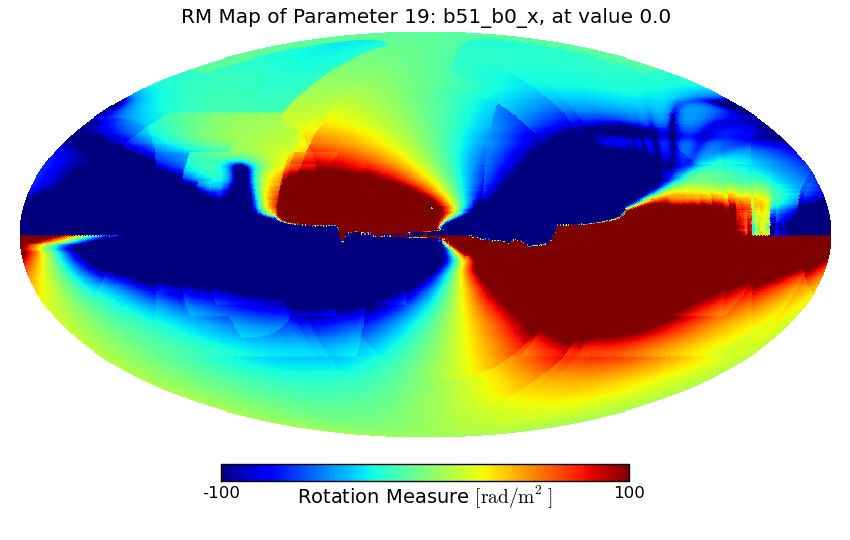}}
	\subfloat{\includegraphics[width=0.33\textwidth]{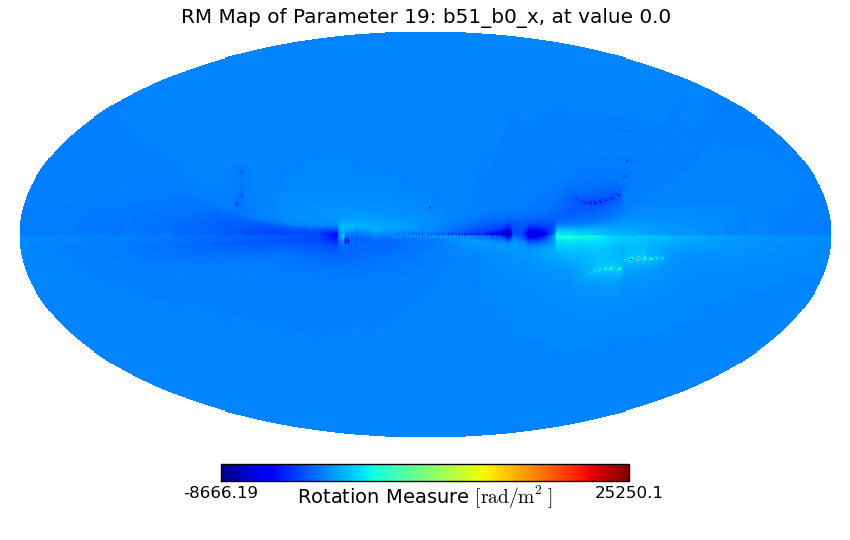}}
	\subfloat{\includegraphics[width=0.33\textwidth]{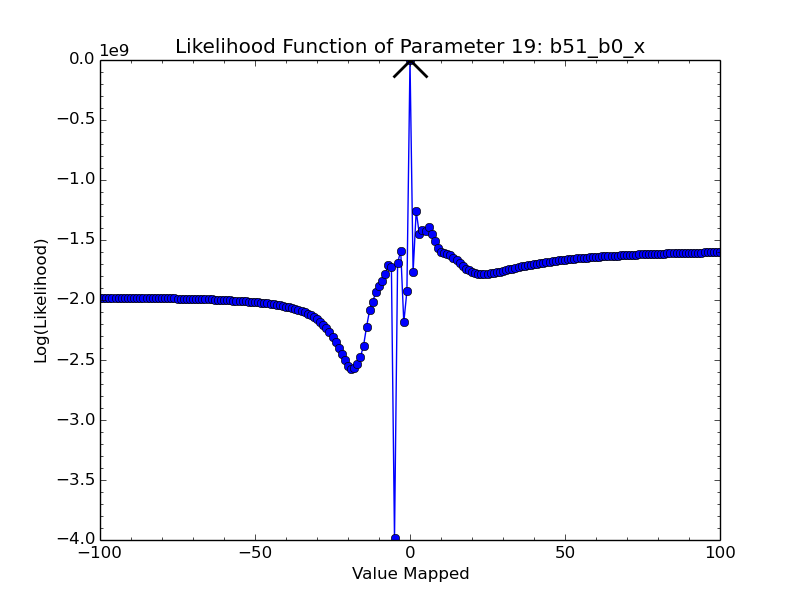}}\\
	\subfloat{\includegraphics[width=0.33\textwidth]{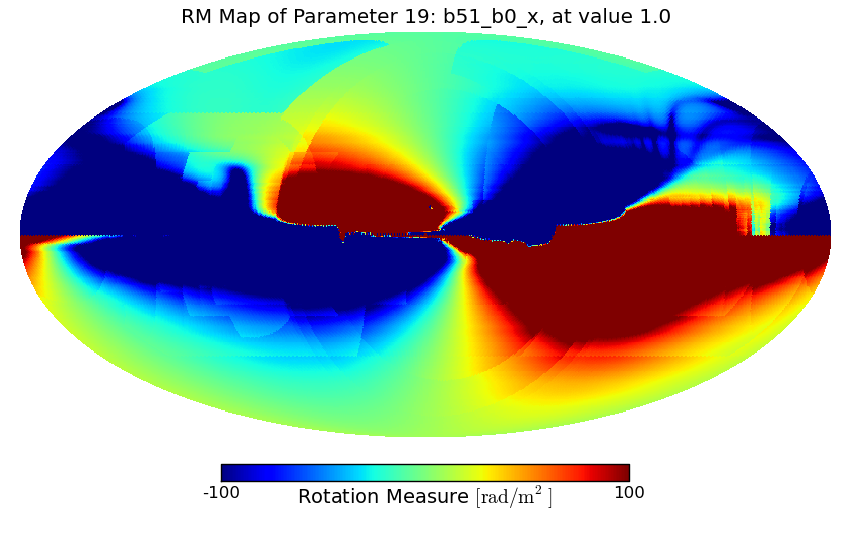}}
	\subfloat{\includegraphics[width=0.33\textwidth]{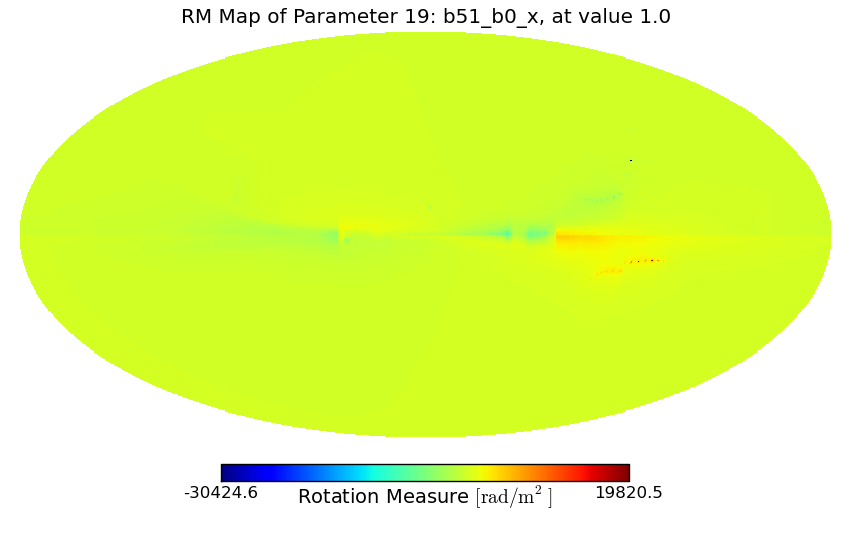}}
	\subfloat{\includegraphics[width=0.33\textwidth]{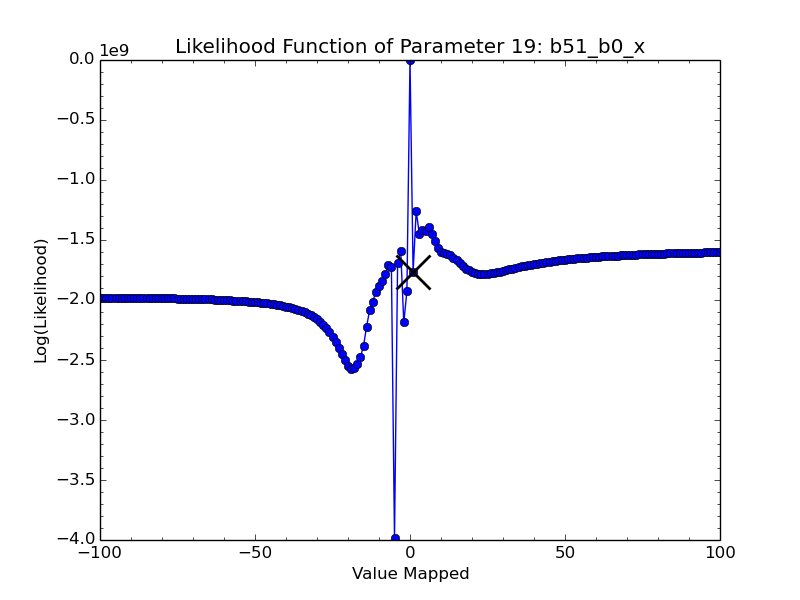}}
	\caption{Figures showing the recreated JF12 RM maps and corresponding likelihood plots for parameter $19$ for three different mapping values (from \textbf{top} to \textbf{bottom}).
	\textbf{Left:} RM maps with their RM values bound to $[-100, 100]\ \mathrm{rad/m^2}$, like the original JF12 RM map.
	\textbf{Center:} RM maps with their RM values unbound.
	\textbf{Right:} Likelihood plots with a cross mark at the corresponding mapping value.}
	\label{fig:RM Movie 19 Old}
\end{center}
\end{figure}
\vref{fig:RM Movie 19 Old} shows two RM maps for three points in the likelihood function: One with the color scale bound to the range $x=[-100, 100]\ \mathrm{rad/m^2}$; and one where the color scale runs from the minimum to the maximum RM value.
The color scale boundaries are chosen with these values to match the boundaries in the JF12 RM map (\vref{fig:JF12 RM}).
\ref{fig:RM Movie 19 Old} shows that the bound RM maps barely show any changes at all, while the unbound RM maps rapidly changes the color of the entire map.
An other detail to notice is that the bound maps show many details (including artifacts), while the unbound maps do not.

\begin{figure}[htb!]
\begin{center}
	\subfloat{\includegraphics[width=0.33\textwidth]{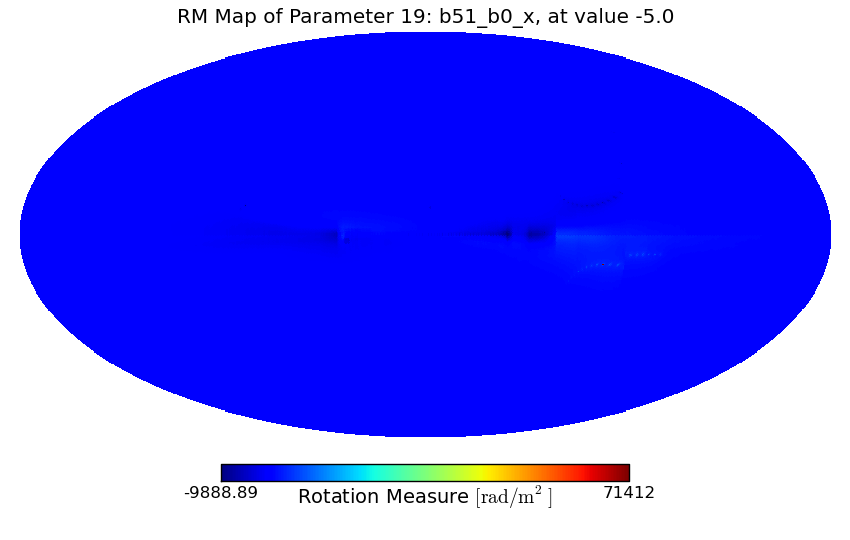}}
	\subfloat{\includegraphics[width=0.33\textwidth]{observable_maps_individual_likelihood/1_1_128_100_35_-/rm_map_p19_v1.png}}
	\caption{Unbound RM maps showing the highest upper bound and lowest lower bound the RM map has taken in the likelihood calculations.
	Both extrema are caused by the same single data point in the map.}
	\label{fig:Data Point Color Change}
\end{center}
\end{figure}

This can be explained when one takes note of the upper and lower bounds of the RM-values in the unbound maps: They vary quite heavily.
This is most likely caused by a single data point which, after some numerical analysis, is discovered to exist.
Its effect can be seen very clearly in \vref{fig:Data Point Color Change}.

Although this single data point is the most extreme in changing its value, removing it from the map would not really be an improvement: The RM map contains many of these rapidly changing data points.
While these data points are most likely numerical artifacts and do not show up in the bound maps, they are still taken into account when calculating the likelihood.
This can mess up any sampling method quite badly, since the same weight is given to extreme, unphysical data points and reasonable, physical data points when using a $\chi^2$--optimization.
A solution for this must be found.

\subsection{Discussion: Part 1}
\label{subsec:Discussion Part 1}
Taking all the results of \vrefrange{subsec:Likelihood Plots Part 1}{subsec:Likelihood Movies Part 1} into account, one can point out a couple of problems:
\begin{itemize}
	\item JF12 contains non-independent parameters;
	\item IMAGINE therefore contains non-independent modules;
	\item Certain parameter sets can give data points that are unphysical (eg. RM-values of $70,000\ \mathrm{rad/m^2}$);
	\item Behavior of likelihood functions of certain parameters cannot be explained (eg. parameter $19$).
\end{itemize}
The first three bullet points affect the efficiency at which a sampler can perform.
As was shown in \vref{fig:Part 1 no random}, disabling the calculation of the random magnetic field renders $13$ random field parameters obsolete.
Although this in itself is not really a problem, the fact that a single parameter influences the behavior of $13$ other parameters, is something unwanted.
If a sampler does not know that this dependency exists, it can easily get stuck in its calculations (trying to find the maximum in a straight horizontal line will result in an infinite loop).
Since the parameter in question is a parameter in Hammurabi that controls the usage of a random magnetic field, this also results in a dependency between the modules in IMAGINE.\footnote{In Hammurabi, the parameter is called \textit{B\_field\_do\_random}.}

The fourth bullet point can be an indication that something numerically is going wrong or is not correctly understood.
Some analysis of the Hammurabi code revealed that there are actually two different JF12 model implementations in the code: Model $7$ and model $51$.
Looking into the differences between these two implementations might say something about the behavior found in the likelihood plot of parameter $19$.

\subsection{Model 7 \& Model 51}
\label{subsec:Model 7 and Model 51}
After taking a closer look at the Hammurabi code, it seems that there are indeed two different implementations of the JF12 model.
One is called model $7$ and is part of the main series of models that can be found in the Hammurabi code.
The other is called model $51$ and is apparently an implementation that is only used for testing purposes.

Everybody in the IMAGINE collaboration that is working with coding (including myself), thought that model $51$ was the correct model to use.
The simple reason for this is that all JF12 parameters are labeled with the prefix \textit{b51}, implying that the original implementation of JF12 must have been model $51$.
We later discovered that this is not true and that model $7$ is the true implementation of JF12.

Looking closely at the different implementation of both models reveals something very interesting: Model $51$ does not have the random field component of the JF12 model implemented.
Model $51$ actually has an approximation of the random (both isotropic and anisotropic random) component coded.
What this means is that the random component of the JF12 model in model $51$ does not behave itself like one would expect from a random field.
Taking the information about Faraday rotation from \vref{subsubsec:Faraday Rotation} into account, one would expect a random field to not have any influence on the outcome of the RM maps due to the RM being a line-of-sight integrated variable.

The random component of JF12 in model $51$ is described as an average of the random field component, which would not cause the weird discontinuities found in \vref{subfig:Parameter 19 do random Part 1} as it still behaves like a random field.
However, some in-depth inspections of the code shows us that this is actually not the case: The approximation of the average of the random field does not behave itself in the same way as a true average would do.
This can then immediately explain the discontinuities in \ref{subfig:Parameter 19 do random Part 1}: The random field component is not working as expected and can thus influence parameters it should not influence.

Model $7$ on the other hand, has the complete random field component of the JF12 model implemented correctly.
In the JF12 model, the isotropic random field component has a $13$-parameter description and is an 'accurate' representation of the isotropic component of the magnetic field of the Milky Way.
However, the JF12 model simply uses two multiplicative factors to describe the anisotropic component of the magnetic field (one that allows scaling with the regular component and the other with the isotropic random component). 
Since the anisotropic component is 'less' important than the isotropic component and because it scales with the isotropic component itself (showing the same behavior), model $7$ behaves a lot more natural than model $51$.

However, when using the correct implementation of the random field component, there is one other thing that should be kept in mind: Do we want to use an average realization or a true realization of the random field?

\subsection{Testing use\_B\_analytic}
\label{subsec:Testing use-B-analytic}
When using the Hammurabi model $7$ instead of model $51$ (see \vref{subsec:Model 7 and Model 51}), one has to deal with a parameter that is called \textit{use\_B\_analytic}.
This parameter is said to specify how the isotropic random fields of a certain model work, but there is no real documentation about the specific workings of this parameter.
For this reason, some research has to be done on how exactly this parameter works such that it can be used in the correct way for the second part of sampler testing in \vref{subsec:Likelihood Plots Part 2}.

\begin{figure}[htbp!]
\begin{center}
	\subfloat[Random = F, Stokes = F]{\includegraphics[width=0.5\textwidth]{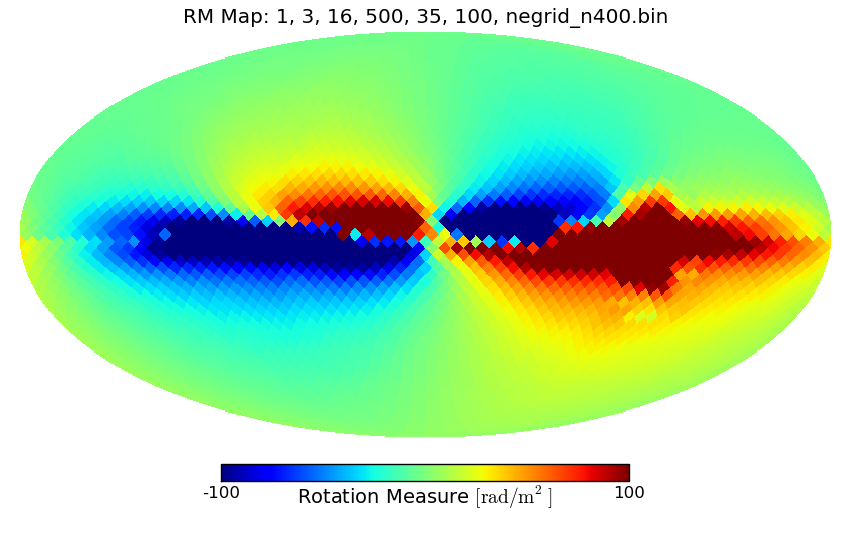}}
	\subfloat[Random = T, Stokes = F]{\includegraphics[width=0.5\textwidth]{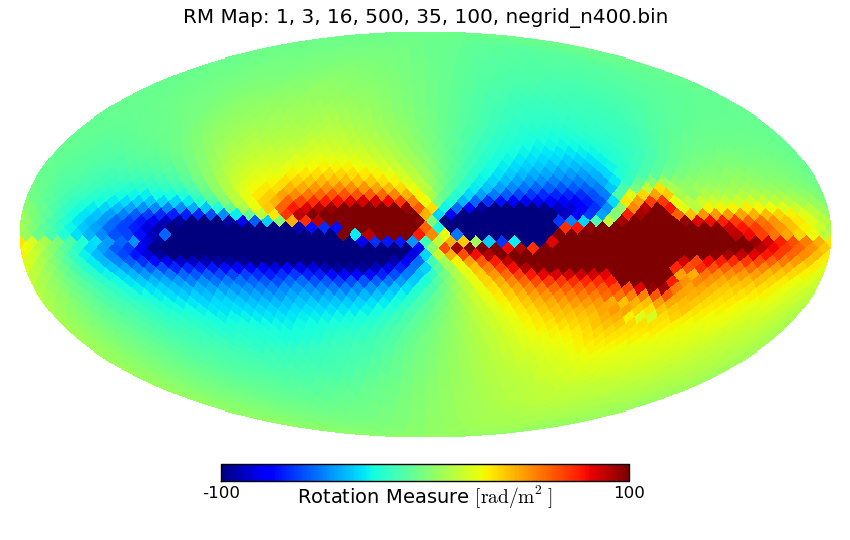}}
\caption{The two different figures obtained when \textit{use\_B\_analytic} is set to True.
The subcaption of the figures states whether or not the calculation of the random field or the Stokes IQU-parameters was included.}
\label{fig:RM maps do analytic}
\end{center}
\end{figure}

To start off, one might want to check what happens if one turns the parameter on and creates some RM maps with it.
The resulting maps can be found in \vref{fig:RM maps do analytic}.
The figure shows that using an isotropic random field does not change anything about the produced RM map if the \textit{use\_B\_analytic} parameter is set to True.
Hammurabi also fails to take the Stokes IQU-parameters (dust and synchrotron) into account in the calculation.
This might be caused by the way that RM maps and Stokes maps are influenced by the presence of a random field: the RM map should not change under the presence of a random field, because the contribution of this field to the RM will average out.
The Stokes maps however, will be influenced by a random field and should thus change.
This gives the idea that \textit{use\_B\_analytic} uses an average realization of the isotropic random field which is perfect for calculating RM maps, but cannot be used for Stokes maps.

\begin{figure}[phtb!]
\begin{center}
	\subfloat[Random = F, Stokes = F]{\includegraphics[width=0.32\textwidth]{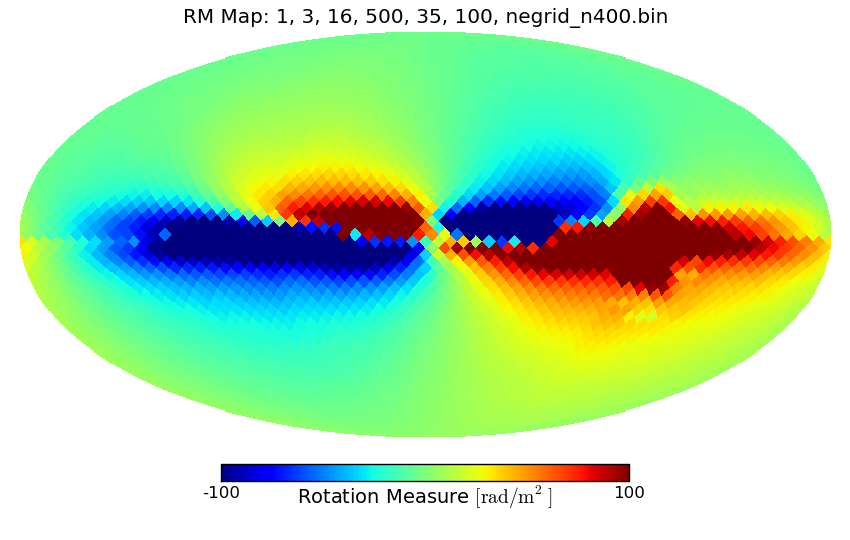}}
	\subfloat[Random = F, Stokes = T]{\includegraphics[width=0.32\textwidth]{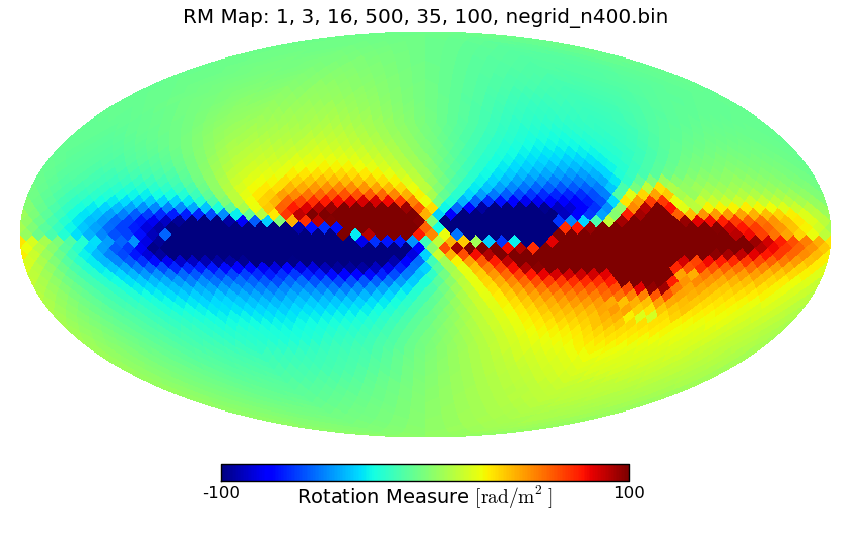}}\\
	\subfloat[Random = T, Stokes = F]{\includegraphics[width=0.32\textwidth]{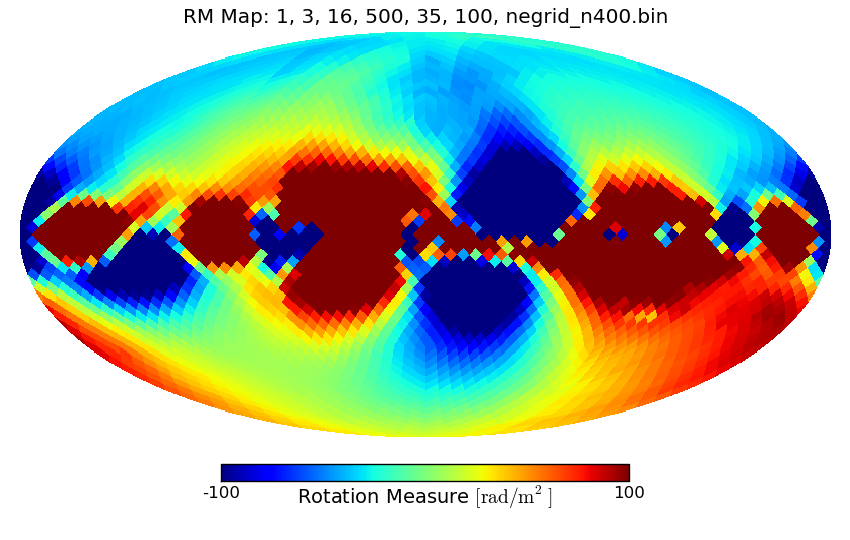}}
	\subfloat[Random = T, Stokes = T]{\includegraphics[width=0.32\textwidth]{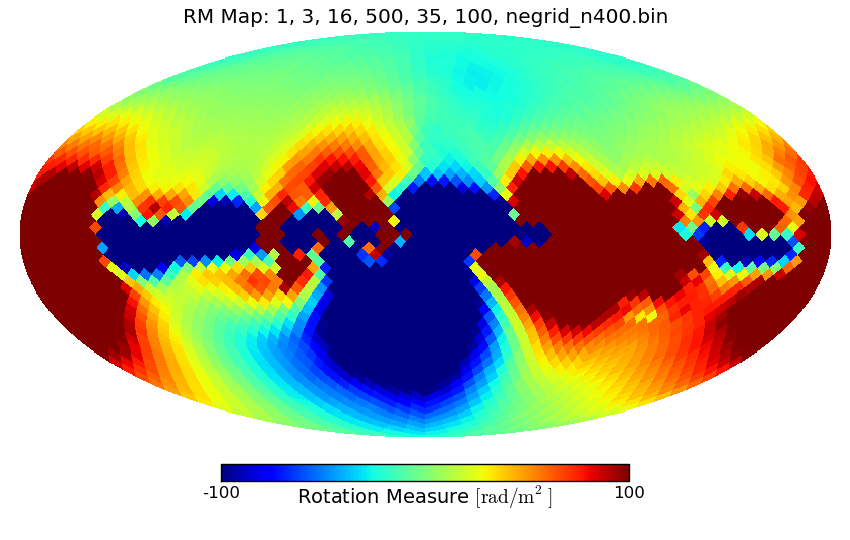}}
\caption{The four different figures obtained when \textit{use\_B\_analytic} is set to False.
These were created in order to check if the Stokes parameters might influence the outcome of the RM maps.
The subcaption of the figures states whether or not the calculation of the random field or the Stokes IQU-parameters was included.}
\label{fig:RM maps no analytic}
\end{center}
\end{figure}
In \vref{fig:RM maps no analytic}, the upper figures show that taking the Stokes parameters into account in the calculation does not influence the produced RM map, which acts as a consistency check.
The lower figures however show that if \textit{use\_B\_analytic} is not used, the produced RM map changes.
Since the upper figures already show that the Stokes parameters do not influence the produced RM map, this most likely means that both figures used a different realization of the isotropic random field.

Although it is usually said that a random field (isotropic or anisotropic) has no influence on the RM, this is actually not true.
If the average of the random field is taken, then no influence on the RM map will be seen.
This is usually taken for granted, because it makes computations easier.
However, the Milky Way does not have an average random field: it has a single realization of the random field and thus the random field influences the produced RM map.

Therefore, if all of this is true, then it means that taking the average RM map of a high number of RM maps using different realizations of the isotropic random field, should produce the figures found in \vref{fig:RM maps do analytic}.

\begin{figure}[phtb!]
\begin{center}
	\subfloat[use\_B\_analytic = F]{\includegraphics[width=0.32\textwidth]{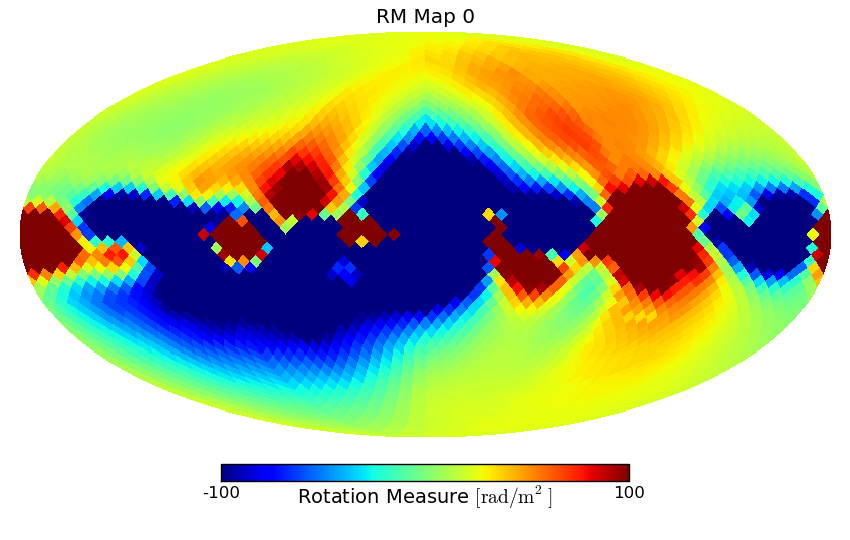}}
	\subfloat[use\_B\_analytic = F]{\includegraphics[width=0.32\textwidth]{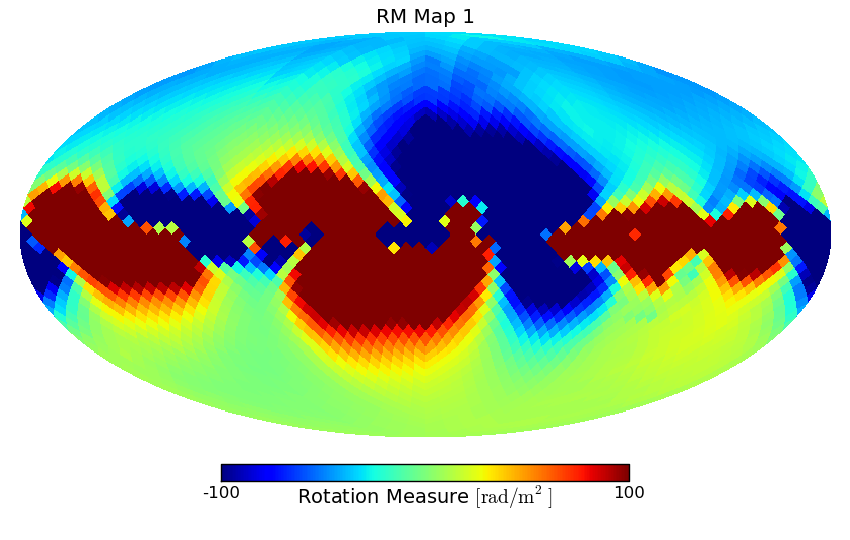}}\\
	\subfloat[use\_B\_analytic RM map]{\includegraphics[width=0.32\textwidth]{use_b_analytic_tests/use_b_analyticT_b_field_interpF/rm_map_1_3_16_500_35_100_negrid_n400_bin_randomT_stokesF.png}}
	\subfloat[Average RM map]{\includegraphics[width=0.32\textwidth]{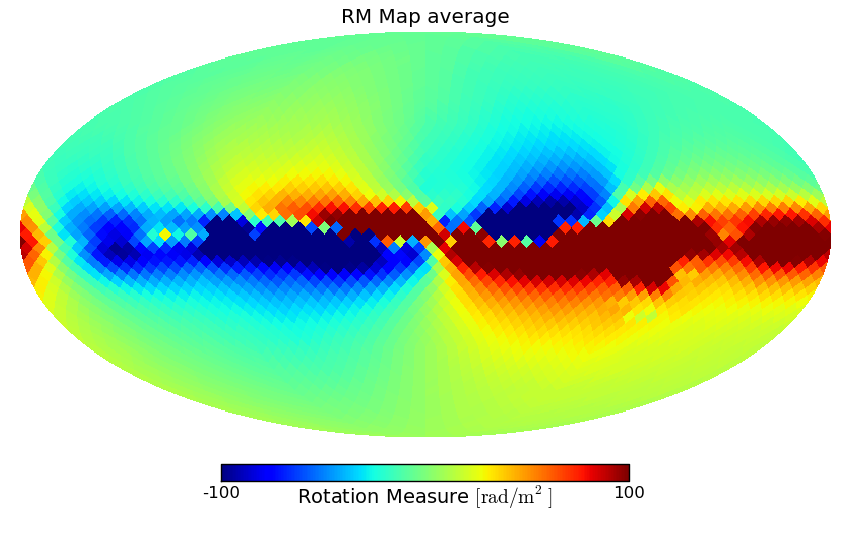}}
\caption{\textbf{Top:} Two examples of RM maps using different realizations of the isotropic random field.
\textbf{Bottom:} Left figure is the RM map with \textit{use\_B\_analytic} set to True (\ref{fig:RM maps do analytic}) and the right figure shows the average RM map of $100$ RM maps using different realizations of the isotropic random field.
All figures had the isotropic random field enabled and the inclusion of the Stokes parameters into the calculation disabled.}
\label{fig:RM maps average}
\end{center}
\end{figure}
Taking $100$ different realizations of the isotropic random field for calculating RM maps should be enough to produce an average RM map that gets close to the real average.
This can be seen in \vref{fig:RM maps average} in the bottom panel: The right figure already shows a lot of similarities to the left figure.
This proves that the parameter called \textit{use\_B\_analytic} controls whether or not an average realization (if set to True) or a true realization (if set to False) of the isotropic random field is used for making the RM maps.
Since the Milky Way itself only has a single realization of the random field, \textit{use\_B\_analytic} will be set to \textit{False} for the next tests performed in \ref{subsec:Likelihood Plots Part 2}.

\subsection{NE2001 or Gaensler?}
\label{subsec:NE2001 or Gaensler?}
\begin{figure}[htb!]
\begin{center}
	\subfloat[Recreated RM map with NE2001]{\includegraphics[width=0.33\textwidth]{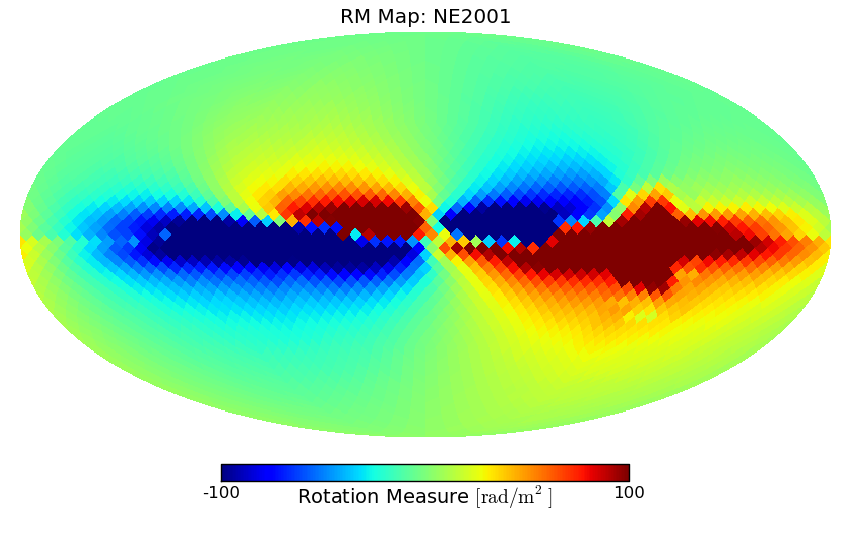}}
	\subfloat[JF12 RM map]{\includegraphics[width=0.33\textwidth]{JF12_RM.png}}
	\subfloat[Recreated RM map with Gaensler]{\includegraphics[width=0.33\textwidth]{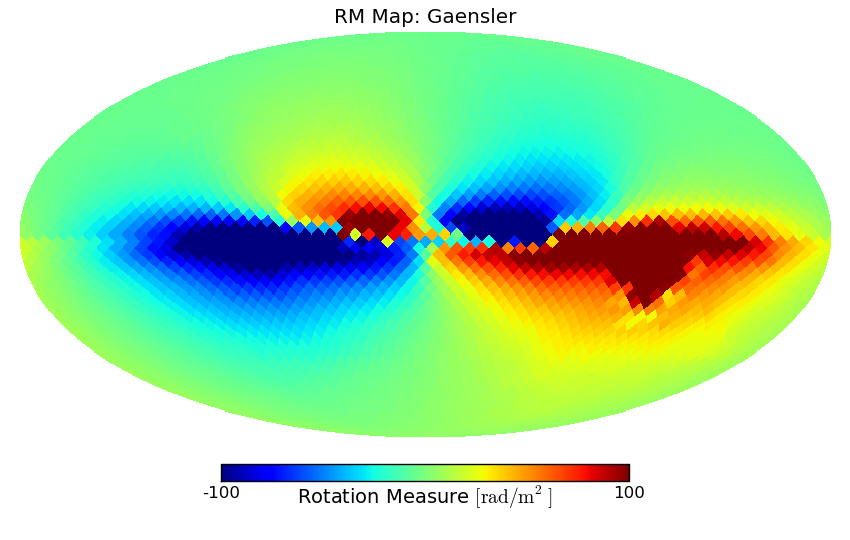}}
	\caption{Comparison of NE2001 and Gaensler with the JF12 RM map.
	\textbf{Left:} RM map created with IMAGINE using JF12 model and NE2001.
	\textbf{Center:} The RM map from the JF12 paper.
	\textbf{Right:} RM map created with IMAGINE using Gaensler.}
	\label{fig:NE2001? Same Resolution}
\end{center}
\end{figure}

\textcite{txt:JF12_regular} state that they use the original NE2001 thermal electron density model discussed in \textcite{txt:NE2001} with the mid-plane density and vertical scale-height modified according to \textcite{txt:Gaensler} (called the 'Gaensler' model from now on).
However, our testing suggests that this might not be the case for their published map as can be seen in \vref{fig:NE2001? Same Resolution}.
Because the data that is in the JF12 RM map itself does not exist anymore and can therefore not be used to numerically compare the maps with each other, this check needs to be done by eye.
Doing so shows something interesting: The NE2001-map shows much more similarities with the JF12 RM map than the Gaensler-map does.
This seems to be an indication that either the JF12 paper is wrong, the wrong model was used for the JF12 RM map by accident or the JF12 RM map used some kind of data filtering that was not reported on in the JF12 paper.
In order to prove this, the authors of the paper have been asked some time ago if they still remember which thermal electron density model was used for their model, to which they negatively replied.
For this reason, it is not possible to prove which of the two models was really used, but it does show that good and accurate documentation is really important when testing models in an automatic pipeline.

\begin{figure}[htb!]
\begin{center}
	\subfloat[Recreated RM map with NE2001]{\includegraphics[width=0.33\textwidth]{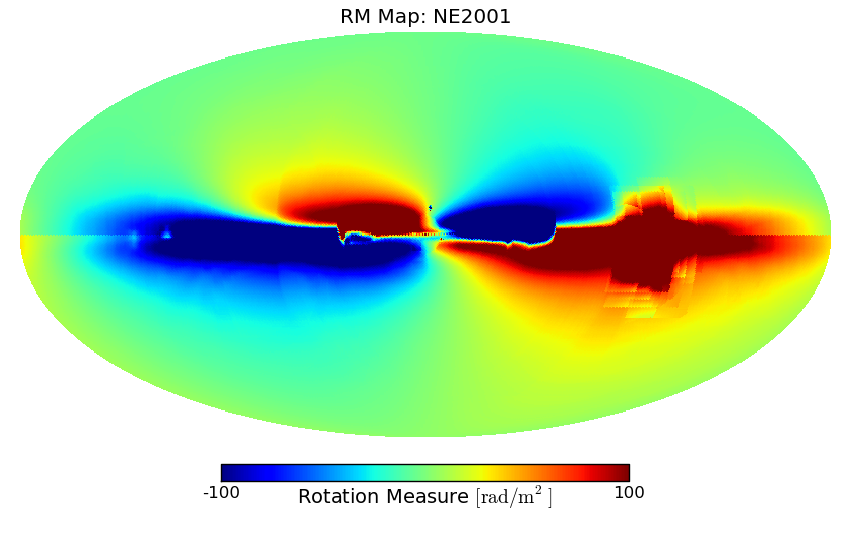}}
	\subfloat[JF12 RM map]{\includegraphics[width=0.33\textwidth]{JF12_RM.png}}
	\subfloat[Recreated RM map with Gaensler]{\includegraphics[width=0.33\textwidth]{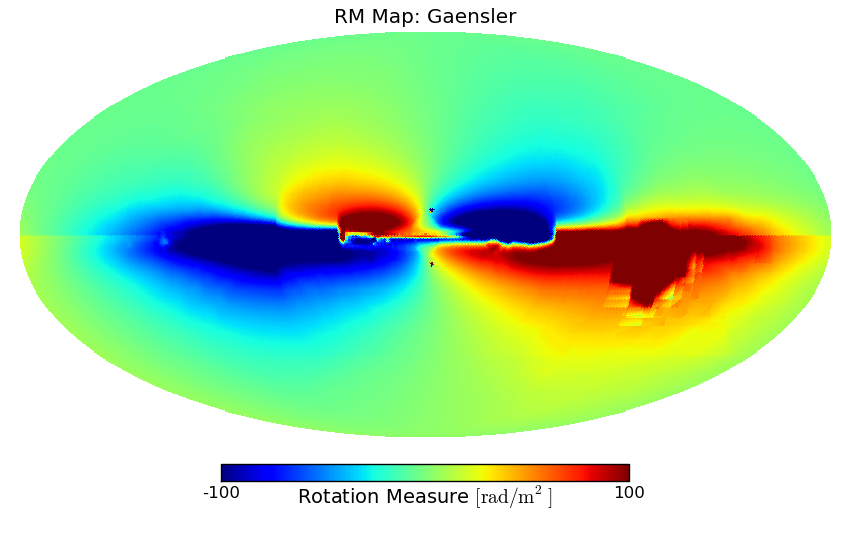}}
	\caption{Same comparison, but now with a much higher resolution.}
	\label{fig:NE2001? Higher Resolution}
\end{center}
\end{figure}

However, papers like \textcite{txt:Schnitzeler} have checked and used the JF12 model as well, but stated that Gaensler was used.
This can maybe be explained by doing the same recreation as done in \ref{fig:NE2001? Same Resolution}, but now with a much higher resolution.
This results in the maps found in \vref{fig:NE2001? Higher Resolution}.

One can argue that both maps have features that compare well to the JF12 RM map: The NE2001-map shows many similarities in the South and not in the North, while the other way around seems to work for the Gaensler-map.
This can however be caused by numerical artifacts or simply the reason that the recreated maps have a much higher resolution than the JF12 RM map itself.
It does show however, that using a higher resolution for the recreation of the maps, can potentially create the illusion that the Gaensler model was used instead of the NE2001 model.
However, since the tests with the same resolution showed that neither map looks completely the same as the JF12 RM map, it is very likely that some kind of filtering was used on the RM map in the JF12 paper, making it impossible to be certain about the thermal electron density model that was used.

\subsection{Likelihood Plots: Part 2}
\label{subsec:Likelihood Plots Part 2}
\begin{figure}[htb!]
\begin{center}
	\subfloat{\label{subfig:Parameter 19 no random}\includegraphics[width=0.33\textwidth]{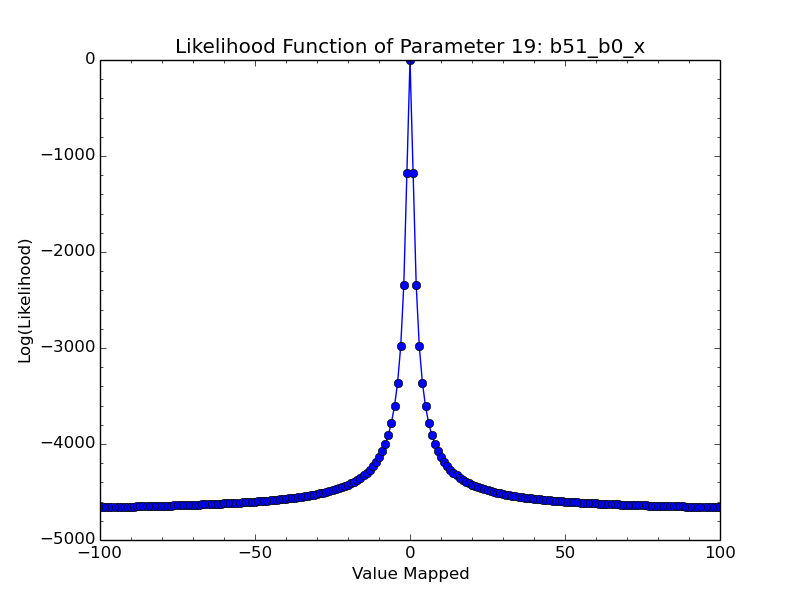}}
	\subfloat{\label{subfig:parameter 20 3 sigma}\includegraphics[width=0.33\textwidth]{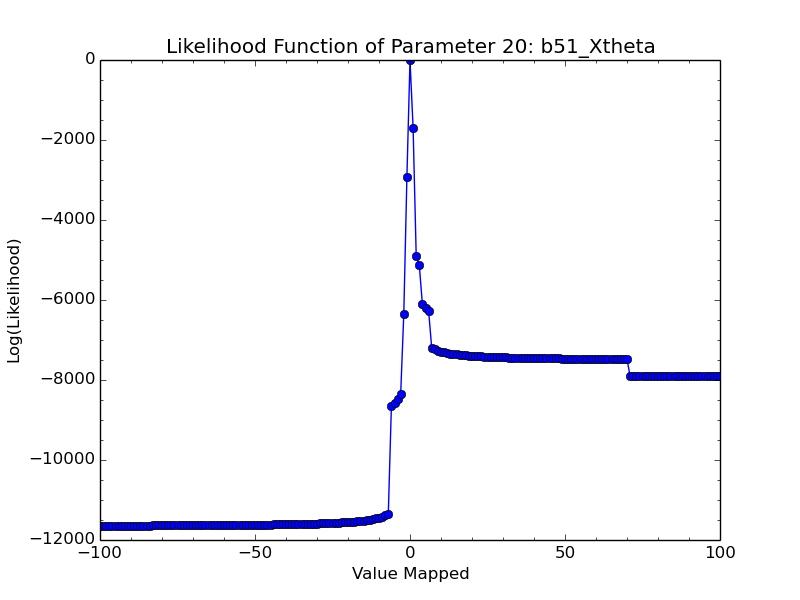}}\\
	\subfloat{\label{subfig:Parameter 28 no random}\includegraphics[width=0.33\textwidth]{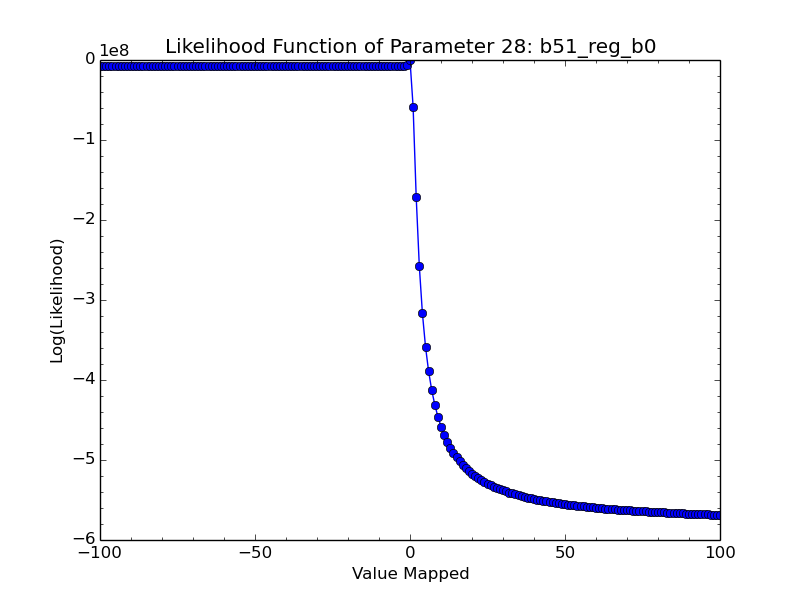}}
	\subfloat{\label{subfig:Parameter 29 no random}\includegraphics[width=0.33\textwidth]{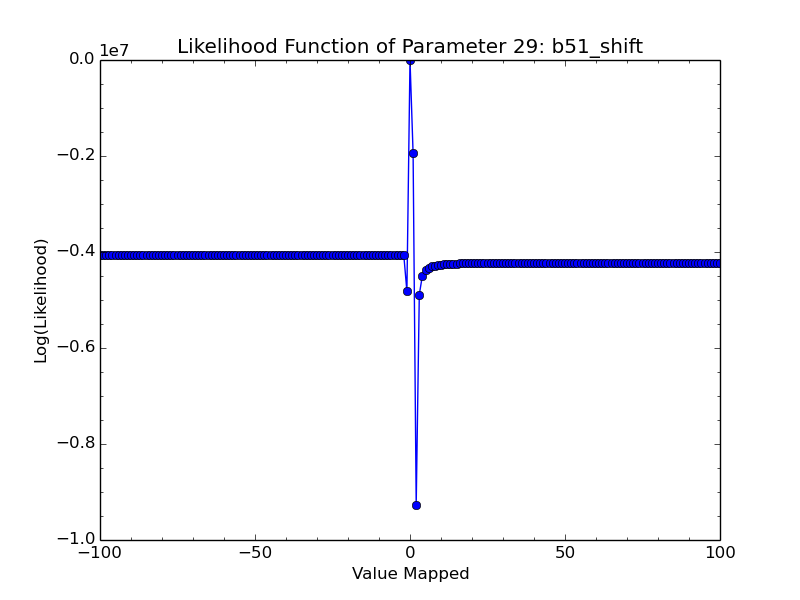}}
	\caption{Likelihoods plots of parameter $19$ (giving the strength of the regular field of the X-halo), parameter $20$ (giving the elevation angle at $z=0$ of the X halo), parameter $28$ (custom parameter scaling the strength of regular field up or down) and parameter $29$ (custom parameter shifting the locations where the arm segments cross the -x axis by a multiplicative amount) of the JF12 model with no random field enabled and usage of Hammurabi model $7$.}
	\label{fig:Part 2 no random}
\end{center}
\end{figure}

The initially conducted tests in the previous sections proved that model $7$ in Hammurabi is a more accurate representation of the JF12 model than model $51$.
Therefore, one needs to check if using model $7$ instead of $51$ also solves some of the other problems that the tests have shown.
Doing so for parameters $19$, $20$, $28$ and $29$ gives the likelihood plots as shown in \vref{fig:Part 2 no random}.
The likelihood plot for parameter $19$ behaves the same as it did in \vref{fig:Part 1 no random}, which is an indication that the regular field still works the same.
Parameter $20$ however, has a likelihood plot that looks semi-continuous: It would have been continuous, if not for the abrupt jumps it shows multiple times throughout the figure.
According to \textcite{txt:JF12_regular}, this parameter sets the elevation angle of the X-halo at the mid-plane ($z=0$) and a galactocentric radius of $4.8\ \mathrm{kpc}$ (default value) or more.
In the conducted tests, it takes values between $46\degree$ and $52\degree$.
This description of the parameter does not sound like something that should behave in a discontinuous way.
However, as stated in \vref{subsec:Testing Details}, all plots and maps are made by using the carrier-mapper.
If a mapping range of $x=[-100, 100]$ is chosen for a $3\sigma$-range, then $x=[-1, 1]$ will already cover the $1.5\sigma$-range.

\begin{figure}[htb!]
\begin{center}
	\includegraphics[width=0.33\textwidth]{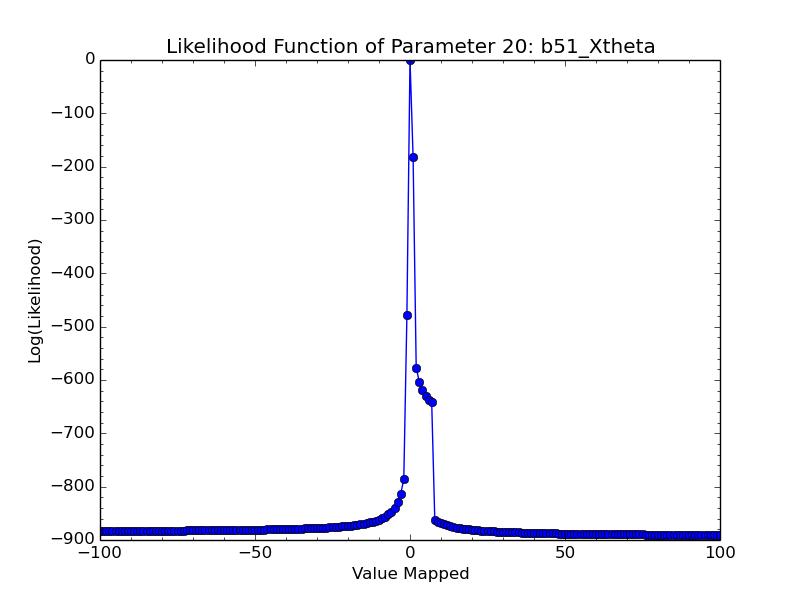}
	\caption{Likelihoods plot of parameter $20$ (giving the elevation angle at $z=0$ of the X halo) of the JF12 model with no random field enabled, usage of Hammurabi model $7$ and a $1\sigma$-range mapped.}
	\label{fig:Part 2 no random 1 sigma}
\end{center}
\end{figure}

Therefore, it might be a good idea to take a look at the same likelihood plot for parameter $20$, but now where a $1\sigma$-range is mapped instead of a $3\sigma$-range.
This plot can be found in \vref{fig:Part 2 no random 1 sigma}.
Note that the $1\sigma$-range is already covered in the mapping range $x=[-0.577, 0.577]$ in \ref{subfig:parameter 20 3 sigma} due to the non-linear scale of the x-axis.
In this plot, one can see that the discontinuity has decreased quite heavily, but still shows up in quite the same fashion as it did in \ref{subfig:parameter 20 3 sigma}.
This plot also shows that although a central peak might look continuous, it can still have discontinuous behavior due to the heavily non-linear scaling of the x-axis.\footnote{Checking the $1\sigma$-range plots for all other parameters showed nothing new.}
This discontinuity cannot currently be explained and might need some further investigating in the future.

The likelihood plot belonging to parameter $29$ in \vref{fig:Part 2 no random}, also shows discontinuous behavior near the default value.
This parameter was originally not in the JF12 model, but was manually added by the current developer of Hammurabi.\footnote{Tess Jaffe, \url{tess.jaffe@nasa.gov}}
Parameter $29$ shifts the locations where the arm segments cross the -x axis by a certain multiplicative amount, that is given in $\mathrm{kpc}$.
Default is to not do any shifting, while moving away from the default value makes the shift-factor be in the range $[0, 10]$.\footnote{Note that this parameter does not have an error, due to it being a shift-factor and not an actual physical parameter. Therefore, checking the $1\sigma$-range plots will not make a difference.}

One could maybe expect some numerical discontinuity far away from the default value, since shifting the spiral arms affects the whole model at once. 
However, this does not happen.
The only discontinuity can be found in the range $x=[-2, +4]$, which gives the values $[0.017, 7.636]$ to parameter $29$.
This is quite a big range due to the way the carrier mapper works, but it does not really explain why the discontinuity exists.
Especially because this parameter does not influence the workings of the model at all.

The likelihood plot of parameter $28$ shows only continuous behavior.
This parameter can scale the regular magnetic field strength up or down, depending on its value.
Interestingly enough, it shows that it is much more probable that the regular magnetic field strength needs to be scaled down than up.
In contrary to the behaviors of parameter $20$ and parameter $29$, this parameter has continuous behavior despite being a parameter that influences the whole model.\footnote{This proves to me that not all global parameters automatically have discontinuous behavior.}

\begin{figure}[htb!]
\begin{center}
	\subfloat{\label{subfig:Parameter 19 do random}\includegraphics[width=0.33\textwidth]{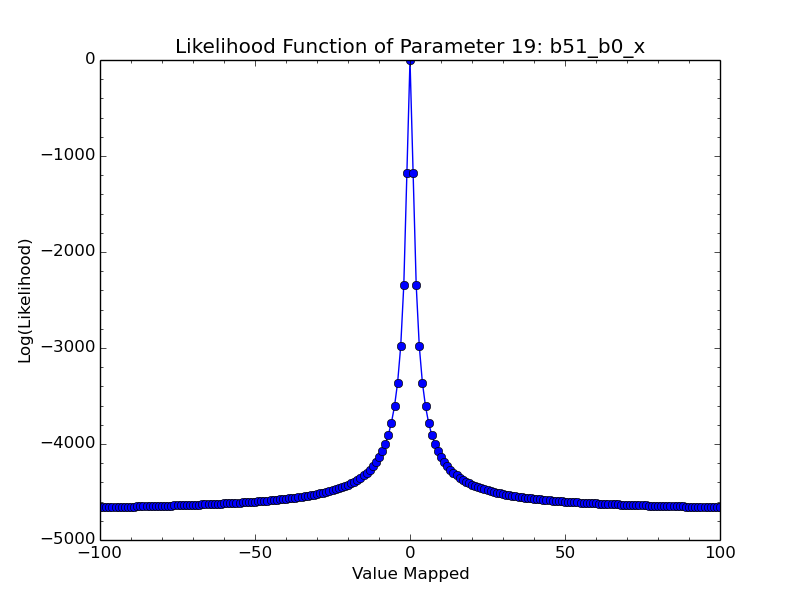}}
	\subfloat{\includegraphics[width=0.33\textwidth]{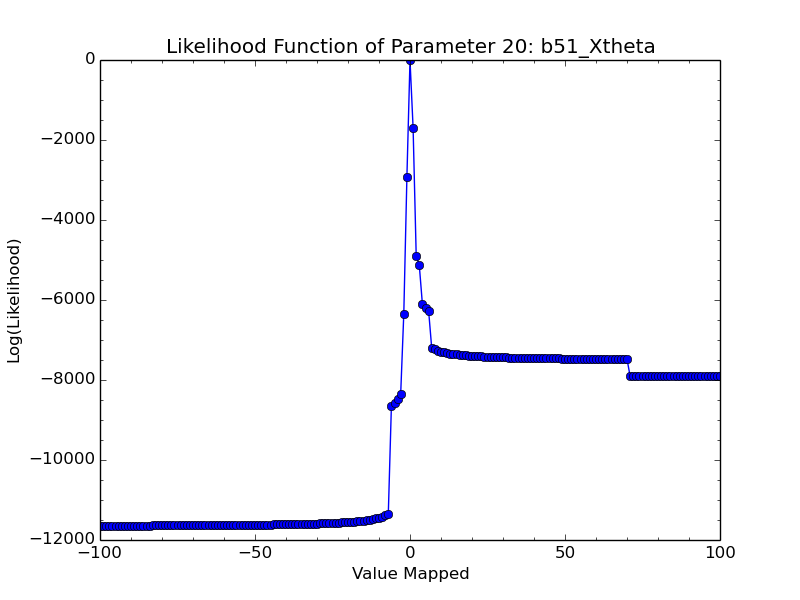}}\\
	\subfloat{\includegraphics[width=0.33\textwidth]{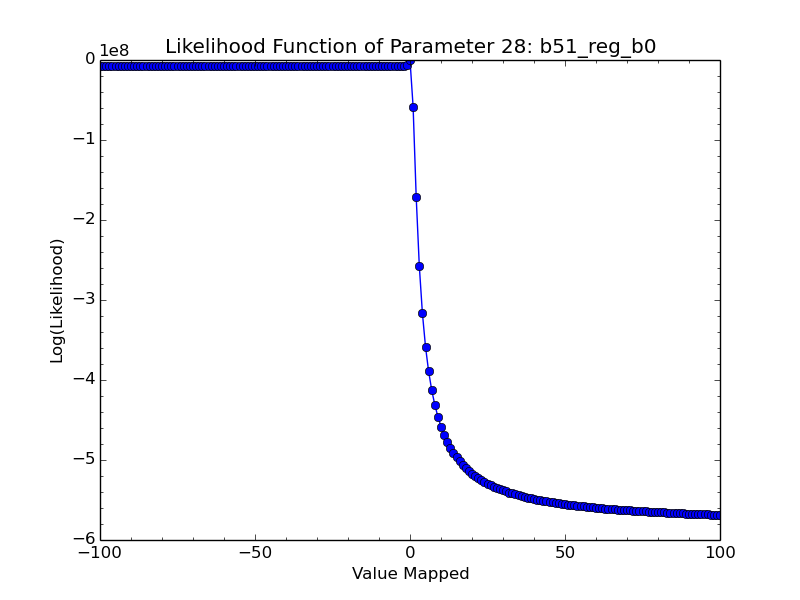}}
	\subfloat{\label{subfig:Parameter 29 do random}\includegraphics[width=0.33\textwidth]{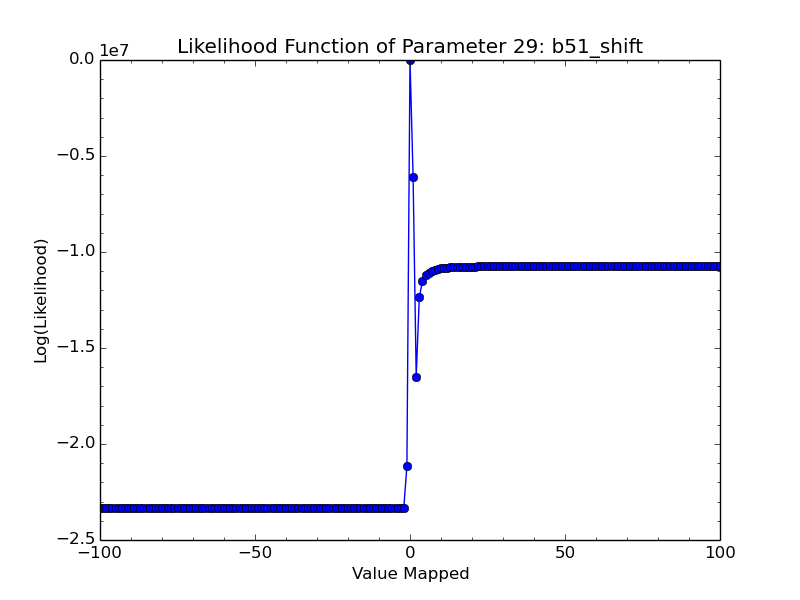}}
	\caption{Likelihoods plots of parameter $19$ (giving the strength of the regular field of the X-halo), parameter $20$ (giving the elevation angle at $z=0$ of the X halo), parameter $28$ (custom parameter scaling the strength of regular field up or down) and parameter $29$ (custom parameter shifting the locations where the arm segments cross the -x axis by a multiplicative amount) of the JF12 model with a single realization of the random field enabled and usage of Hammurabi model $7$.}
	\label{fig:Part 2 do random}
\end{center}
\end{figure}

Maybe turning on the random magnetic field will give some answers about the discontinuities found in parameters $20$ and $29$.
Note that the coded model $7$ is now used instead of $51$, which can use both a single realization and an average realization of the random field.
Since using a random realization of the random field makes no sense when simulating Faraday rotation, a single realization will be used.
The resulting plots are displayed in \vref{fig:Part 2 do random}.
This time, parameter $19$ behaves exactly as predicted: The random magnetic field component does not influence it.\footnote{Numerically comparing \ref{subfig:Parameter 19 no random} with \ref{subfig:Parameter 19 do random} shows that both plots are indeed identical.}

The plot for parameter $20$ also seems to not being influenced by the random magnetic field component.\footnote{Which numerically comparing shows as well.}
However, since all likelihood plots of parameter $20$ show this discontinuous behavior, one can argue that the regular magnetic field is causing these effects, but that is highly doubtful.
If this is the case, then this needs to be investigated further.
I have not been capable of finding any explanation for why this happens.

Parameter $28$'s plot is still exactly the same, which again is expected of a parameter that only influences the regular magnetic field component.
The likelihood plot of parameter $29$ is most likely the weirdest one: It still shows discontinuous behavior, but it is different from the one seen in \ref{subfig:Parameter 29 no random}.
The positive mapped values seem to behave roughly the same as before (although still giving a different likelihood value), but the negative values now give a lower likelihood than the positive values, unlike before.
This basically means that turning on the random field makes it a bit more likely that the spiral arms need to be shifted positively than negatively.
Since the other way around is true for \ref{subfig:Parameter 29 no random}, something strange must be going on in these parameters that affect the model globally.
For this, I do not have an explanation yet either.

\subsection{Discussion: Part 2}
\label{subsec:Discussion Part 2}
With the second part of the testing done, a few things are now clear:
\begin{itemize}
	\item Correct description of the random field component is much more complex than the regular field component;
	\item Discontinuous likelihood functions can and most likely will exist in GMF models, which require a nested sampler;
	\item It is practically unavoidable to have parameters in a GMF model that are dependent on each other, signifying the usage of a Gibbs sampling method inside the nested sampler;
	\item It is better to write a separate GMF generator module for the IMAGINE pipeline instead of using the internal GMF generator of Hammurabi to avoid any dependencies;
	\item Both the sampler and the GMF generator need to be capable of handling an optimization request over only a part of the model parameters, not all of them;
	\item The simple $\chi^2$-optimization might be too simple and needs something more sophisticated;
	\item The carrier mapper has its weak points: its emphasis on generating points far away from the default value might be too strong.
\end{itemize}

That the IMAGINE pipeline requires a nested sampler can be seen very clearly in \vref{fig:Part 2 no random 1 sigma}.
This parameter shows discontinuity in the likelihood function.
Although this discontinuity is not necessarily as bad as the ones found in \vref{subfig:Parameter 29 no random} and \vref{subfig:Parameter 29 do random}, parameter $20$ is much more important for the workings of the model than parameter $29$.\footnote{Parameter $29$ is simply a custom shifting factor with unexplainable discontinuity. The discontinuity is most likely artificial and thus not necessarily important for the sampler.}

\begin{figure}[htb!]
\begin{center}
	\includegraphics[width=\textwidth]{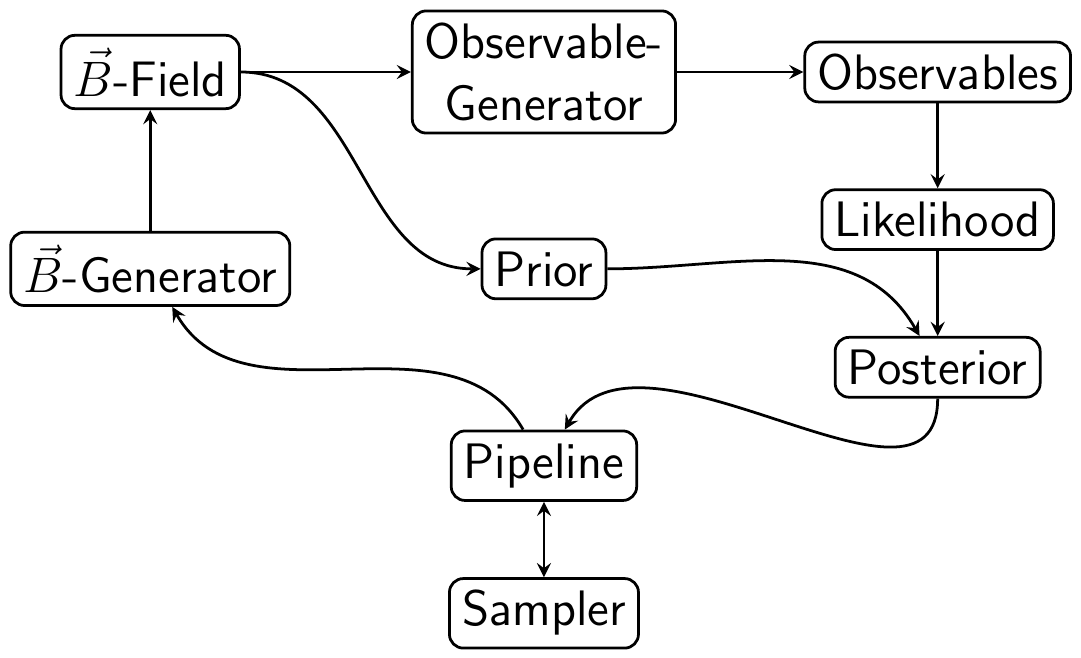}
	\caption{An updated schematic overview of the IMAGINE pipeline.
	Credits to Theo Steininger.}
	\label{fig:Updated IMAGINE pipeline}
\end{center}
\end{figure}

Because of these points, some of them have already been implemented in the IMAGINE pipeline during the research of this thesis.
Therefore, a new pipeline chart was also created in order to reflect the changes that have been made to the pipeline.
This new chart can be seen in \vref{fig:Updated IMAGINE pipeline}.

The biggest change that have been made is the separation of the GMF generator module and the observable generator module.
These two modules are no longer taken care of by the Hammurabi code, but the GMF generator module has gained its own module.
Hammurabi is capable of taking an external GMF-map as its input instead of a coded model, so this can work perfectly fine.

Another big change that came out of my testing, is the necessity of having an option to only optimize a part of the parameters of a model.
This came out of the dependency between the random field parameter in Hammurabi and the $13$ isotropic random field parameters in JF12.
Although these two modules have now been separated, it still made clear that it is very useful to be capable of only optimizing certain components of a model.
Now, one can tell the pipeline that one for example only wants to optimize the regular field component of a model, as long as it is known to the GMF generator and the sampler which parameters it has to choose for this.

A third important change that came out, is that a simple $\chi^2$-optimization is not nearly good enough for testing models.
Especially in \vref{fig:RM Movie 19 Old}, this can be seen very clearly.
Although we now know that most of the extreme RM values in the maps was caused by usage of model $7$ instead of $51$, it was still a 'valid' model.
Therefore, it is possible that future models can also give values that make no sense.
When such values are present in the observable maps, they should get a much lower priority than the other values.
Since there is no way of knowing if and how many of these values are present, something more sophisticated is required for calculating the likelihood.

\newpage
\section{Discussion \& Conclusion}
\label{sec:Discussion and Conclusion}
In this thesis, I have shown that in order to guarantee compatibility with all potential GMF models, the IMAGINE pipeline needs several adjustments or improvements.
The most important thing that we have learned from the performed tests, is that GMF models can show discontinuous behavior.
In the case of the JF12 model, I have shown multiple different ways of explaining these discontinuous features that were found in the likelihood plots.
However, not all of them can be explained.
Some of them were explained as being 'numerical' and have been removed, while others were explained as being 'physical' or at least expected due to an other numerical effect (parameter $19$ showing discontinuous behavior was caused by the usage of model $51$).
Since there is no guarantee that every model the IMAGINE pipeline will see, is a perfectly well-parametrized model, it forces us to take care of this by using a nested sampler.

Using the nested sampler also has the additional benefit that it gives the evidence as its main result, and not the posterior.
This automatically makes the tested model future-proof when new models need testing, which arguably is also a nice property to have.
A nested sampler has since already been implemented into the IMAGINE pipeline (given by the Python-package \textsc{PyMultiNest}).

I have also shown that a simple $\chi^2$-optimization is not enough when one starts dealing with models that can actually have flaws.
As long as a model is capable of producing data that physically makes no sense (like RMs with values over $70,000\ \mathrm{rad/m^2}$), the pipeline needs to take care of this by adding a lower priority to the data points that cause them.
This pointed out that a more complex way of calculating the likelihood probability is required, which has since been implemented into the IMAGINE pipeline.

Something that I have shown as well during my tests, is that using code that was written by others can make problems a lot more complex.
Although nobody cannot really be blamed for documenting their code poorly, it does however cause many headaches and incorrect interpretations.
This exact problem can be noted fairly well during my tests with model $7$ and $51$ in Hammurabi.
The Hammurabi code is a nice code with a really useful application, but it has very poor documentation.
Many parameter names in the code do not really describe what they are used for, or actually imply something different (like the fact that all JF12 parameters have the prefix \textit{b51}, implying that model $51$ contains the full JF12 model).
This acts both as a warning to others and to myself, that we always have to take care when using code not written by ourselves.

\newpage
\bibliographystyle{mnras}
\bibliography{References}

\newpage
\clearpage
\pagenumbering{roman}
\labelformat{section}{App.~#1}
\labelformat{subsection}{App.~#1}
\labelformat{subsubsection}{App.~#1}
\begin{appendices}
\section{Derivations}
\label{app:Derivations}
This appendix contains the derivations of certain important equations used in this thesis.
Whenever there is a reference to an equation inside a derivation, it indicates that either a certain part of the derivation equals this equation or that this equation will get substituted into the derivation.

\subsection{\ref{eq:Bayes' Theorem}}
\label{der:Bayes' Theorem}
Starting with \ref{eq:Product Rule}, the transpose version of this equation can be written down:
\begin{eqnarray}
\label{eq:Product Rule Transposed}
\mathrm{prob}(Y,X|I)&=&\mathrm{prob}(Y|X,I)*\mathrm{prob}(X|I).
\end{eqnarray}
Since the statement that '$Y$ and $X$ are both true' is the same as '$X$ and $Y$ are both true', such that $\mathrm{prob}(Y,X|I)=\mathrm{prob}(X,Y|I)$, \ref{eq:Product Rule Transposed} can be equated to \ref{eq:Product Rule}.
This gives
\begin{eqnarray*}
\mathrm{prob}(X|Y,I)*\mathrm{prob}(Y|I)&=&\mathrm{prob}(Y|X,I)*\mathrm{prob}(X|I);\\
\Rightarrow \mathrm{prob}(X|Y,I)&=&\frac{\mathrm{prob}(Y|X,I)*\mathrm{prob}(X|I)}{\mathrm{prob}(Y|I)}.
\end{eqnarray*}

\subsection{\ref{eq:Bayes' Marginalization}}
\label{der:Bayes' Marginalization}
By expanding $\mathrm{prob}(X,Y|I)$ and $\mathrm{prob}(X,\overline{Y}|I)$ with \ref{eq:Product Rule}, one obtains the following:
\begin{eqnarray}
\nonumber
\mathrm{prob}(X,Y|I)&=&\mathrm{prob}(Y,X|I);\\
\label{eq:Product Expansion 1}
&=&\mathrm{prob}(Y|X,I)*\mathrm{prob}(X|I);\\
\label{eq:Product Expansion 2}
\mathrm{prob}(X,\overline{Y}|I)&=&\mathrm{prob}(\overline{Y}|X,I)*\mathrm{prob}(X|I).
\end{eqnarray}
Now, by combining \ref{eq:Product Expansion 1} with \ref{eq:Product Expansion 2}, one can write:
\begin{eqnarray}
\nonumber
\mathrm{prob}(X,Y|I)+\mathrm{prob}(X,\overline{Y}|I)&=&\underbrace{\left(\mathrm{prob}(Y|X,I)+\mathrm{prob}(\overline{Y}|X,I)\right)}_{\ref{eq:Sum Rule}}*\mathrm{prob}(X|I);\\
\label{eq:Standard Marginalization}
\mathrm{prob}(X|I)&=&\mathrm{prob}(X,Y|I)+\mathrm{prob}(X,\overline{Y}|I).
\end{eqnarray}
Stated verbally, \ref{eq:Standard Marginalization} says that the probability that $X$ is true, irrespective of $Y$ being true or not, is equal to the probability that both $X$ and $Y$ are true plus the probability that $X$ is true and $Y$ is false.

\ref{eq:Standard Marginalization} only includes a single proposition $Y$ and its negative counterpart $\overline{Y}$.
Suppose $Y$ is changed to include all possibilities of $Y$, then the proposition $Y$ changes to a set of possibilities: $Y_1, Y_2, ..., Y_M=\{Y_k\}$.
If $\{Y_k\}$ forms a mutually exclusive and exhaustive set of possibilities, meaning that exactly one $Y_k$ must be true and all the others must be false at all times, the \textit{normalization} requirement can be written down:
\begin{eqnarray}
\label{eq:Normalization Requirement}
\sum_{k=1}^{M}\mathrm{prob}(Y_k|X,I)&=&1.
\end{eqnarray}
By using \ref{eq:Product Rule} to prove that
\begin{eqnarray}
\nonumber
\mathrm{prob}(X,Y_k|I)&=&\mathrm{prob}(Y_k|X,I)*\mathrm{prob}(X|I);\\
\label{eq:Product Expansion 3}
\sum_ {k=1}^{M}\mathrm{prob}(X,Y_k|I)&=&\sum_ {k=1}^{M}\mathrm{prob}(Y_k|X,I)*\mathrm{prob}(X|I),
\end{eqnarray}
\ref{eq:Standard Marginalization} can be written in a way to include the set of possibilities $\{Y_k\}$:
\begin{eqnarray}
\nonumber
\mathrm{prob}(X|I)&=&1*\mathrm{prob}(X|I);\\
\nonumber
&\stackrel{\ref{eq:Normalization Requirement}}{=}&\sum_{k=1}^{M}\mathrm{prob}(Y_k|X,I)*\mathrm{prob}(X|I);\\
\label{eq:Summed Marginalization}
&\stackrel{\ref{eq:Product Expansion 3}}{=}&\sum_{k=1}^{M}\mathrm{prob}(X,Y_k|I).
\end{eqnarray}
\ref{eq:Summed Marginalization} is a generalization of \ref{eq:Standard Marginalization}.
The true marginalization equation of \ref{eq:Bayes' Marginalization} can be obtained by using the \textit{continuum} generalization of \ref{eq:Summed Marginalization}.
Instead of assigning a certain amount of values to $Y_k$, an arbitrarily large number of propositions is considered about the range in which $Y_k$ might be.
As long as the intervals are chosen in a contiguous way and cover a big enough range of values, a mutually exclusive and exhaustive set of possibilities will be obtained.
\ref{eq:Bayes' Marginalization} is then just a generalization of \ref{eq:Summed Marginalization} with $M\rightarrow \infty$.
This changes the meaning of $Y_k$ to a value $Y$ that represents the numerical value of a parameter of interest and the integrand $\mathrm{prob}(X,Y|I)$ is technically a density function of probabilities rather than a normal probability (and thus should be called a \textit{probability density function}).
Therefore, it should be denoted by a different symbol like $\mathrm{pdf}(X,Y|I)$, where
\begin{eqnarray*}
\mathrm{pdf}(X,Y=y|I)&=&\lim_{\delta y\rightarrow 0}\frac{\mathrm{prob}(X,y\leq Y<y+\delta y|I)}{\delta y},
\end{eqnarray*}
and the probability that the value of $Y$ lies in a finite range between $y_1$ and $y_2$ (given that $X$ is true) is given by
\begin{eqnarray}
\label{eq:Integrated Marginalization}
\mathrm{prob}(X,y_1\leq Y<y_2|I)&=&\int_{y_1}^{y_2}\mathrm{pdf}(X,Y|I)dY.
\end{eqnarray}
Since the word 'pdf' is commonly used as an abbreviation for \textit{probability distribution function}, which says something about a discrete set of possibilities rather than a continuous set, the word 'prob' is used for anything related to probabilities throughout this thesis.
Thus, by taking the infinite case of \ref{eq:Integrated Marginalization}, one can finally obtain \ref{eq:Bayes' Marginalization}:
\begin{eqnarray*}
\mathrm{prob}(X,-\infty<Y<\infty|I)&=&\int_{-\infty}^{\infty}\mathrm{prob}(X,Y|I)dY;\\
\Rightarrow\mathrm{prob}(X|I)&=&\int_{-\infty}^{\infty}\mathrm{prob}(X,Y|I)dY.
\end{eqnarray*}

\newpage
\section{Abbreviations/Terminology}
\label{app:Abbreviations/Terminology}
This appendix contains a list with all used abbreviations and terminology in this thesis.

\paragraph{Abbreviations}
\begin{itemize}
	\item CR: Cosmic Ray;
	\item GMF: Galactic Magnetic Field;
	\item HMC: Hamiltonian/Hybrid Monte Carlo;
	\item IMAGINE: Interstellar MAGnetic field INference Engine;
	\item ISM: InterStellar Medium;
	\item JF12: The papers by \citet{txt:JF12_regular,txt:JF12_random};
	\item MCMC: Markov Chain Monte Carlo;
	\item MH: Metropolis-Hastings;
	\item PDF: Probability Distribution Function;
	\item RM: Rotation Measure;
	\item SNR: SuperNova Remnant;
	\item UHECR: Ultra-High-Energy Cosmic Ray.
\end{itemize}

\paragraph{Terminology}
\begin{itemize}
	\item \underline{Anisotropic random} magnetic field: The variant of the small-scale magnetic field component that shows variations in magnetic field strength and direction, but not in orientation;
	\item \underline{Isotropic random} magnetic field: The variant of the small-scale magnetic field component that shows variations in magnetic field strength, direction and orientation;
	\item \underline{Random} magnetic field: The general small-scale magnetic field component of the Milky Way;
	\item \underline{Regular} magnetic field: The large-scale magnetic field component of the Milky Way.
\end{itemize}
\end{appendices}
\end{document}